\documentclass[amsmath,amssymb,aps, pra,preprint,nofootinbib]{revtex4-1}
 
\usepackage{graphicx, bbold}
\usepackage{dcolumn,soul,bbold}
\usepackage[table,xcdraw]{xcolor}

\usepackage{color}
\numberwithin{equation}{section}

\usepackage{orcidlink}

\usepackage{hyperref}
\hypersetup{
  colorlinks   = true, 	
  urlcolor     = cyan, 	
  linkcolor    = magenta, 	
  citecolor   = blue 	
}

\begin{document}

\title{Planar Dirac equation with radial contact potentials}

\author{J.T. Lunardi\,\orcidlink{0000-0001-7058-9592}}
\email[]{jttlunardi@uepg.br}
\affiliation{Departamento de Matemática e Estatística, Universidade Estadual de Ponta Grossa, 84320-900, Ponta Grossa, PR, Brazil}
\author{S. Salamanca\,\orcidlink{0000-0003-0151-8373}}
\email[]{sergio.salamanca@uva.es}
\affiliation{Departamento de Física Teórica, Atómica y Óptica, and Laboratory for Disruptive \\ Interdisciplinary Science (LaDIS), Universidad de Valladolid, 47011 Valladolid, Spain}
\author{J. Negro\,\orcidlink{0000-0002-0847-6420}}
\email[]{jnegro@uva.es}
\affiliation{Departamento de Física Teórica, Atómica y Óptica, and Laboratory for Disruptive \\ Interdisciplinary Science (LaDIS), Universidad de Valladolid, 47011 Valladolid, Spain}
\author{L.M. Nieto\,\orcidlink{0000-0002-2849-2647}}
\email[]{luismiguel.nieto.calzada@uva.es}
\affiliation{Departamento de Física Teórica, Atómica y Óptica, and Laboratory for Disruptive \\ Interdisciplinary Science (LaDIS), Universidad de Valladolid, 47011 Valladolid, Spain}

\date{\today}

\begin{abstract}
We investigate the planar Dirac equation with the most general time-independent contact (singular) potential supported on a circumference. Taking advantage of the radial symmetry, the problem is effectively reduced to a one-dimensional one (the radial), and the contact potential is addressed in a mathematically rigorous way using a distributional approach  that was  originally developed to treat point interactions in one dimension,  providing  a physical interpretation for the interaction parameters.  
The  most general contact interaction for this system is obtained in terms of four physical parameters: the strengths of a scalar and the three components of a singular Lorentz vector potential  supported on the circumference. We  then  investigate the bound and scattering  solutions for several choices of the physical parameters, and analyze the confinement properties of the corresponding potentials. 
\end{abstract}

\maketitle


\section{Introduction}
\label{intro}

Much work has recently been done on the role of electric and magnetic fields for technological applications in the confinement of relativistic massive charged particles, especially in two-dimensional systems governed by the Dirac-Weyl equation
\cite{hall04,castro09,martino07,negro09,negro18,portnoi16b,apalkov08,bardarson09,peeters08,Hartmann14,Portnoi17,portnoi11,portnoi12,portnoi19,miserev,ho14}, although it is known that this is not suitable for massless two-dimensional particles due to the existence of a strong Klein tunneling effect \cite{katnelson06,beenakke08}. Another situation where electric fields are able to confine particles is when we deal with a quantum dot with some kind of symmetry, typically radial, that allows for a well-defined angular momentum quantum number.
Confinement of massive particles in two dimensions has also been studied to better understand new topological materials, such as graphene, silicene, germanene, stanene and phosphorene \cite{Cayssol13,Shen17,candido18}, where charges acquire effective mass from spin-orbit interaction and perpendicular electric fields \cite{liu11,KNS25}.
An important consequence of massive particle confinement is that it allows to observe the phenomenon of atomic collapse, which is very difficult to detect in relativistic quantum mechanics but is accessible with Dirac materials \cite{peeters16, peeters18, peeters19, peeters21} due to the existence of the Fermi velocity $v_F<\!<c$ \cite{peeters17}. The search for resonances is also an interesting topic, and in fact confinement in Dirac materials must necessarily include bound and resonance states~\cite{Jakubsky22}.

Although it is clear that the aforementioned technological applications  are extremely interesting, in this work our focus will be on the theoretical analysis of certain two-dimensional systems that are governed by the (2+1)  massive Dirac equation subjected to a type of toy potentials, the singular contact interactions in two dimensions. It is well known that contact or point interactions in one-dimensional relativistic quantum mechanics have been used to model short-range potentials. They are singular interactions whose appropriate treatment needs  the use of special mathematical methods in order to avoid ambiguities (see \cite{frontiers} and references quoted therein). 
The main goal of the present work is to adapt a recent distributional approach, originally developed for the one-dimensional case \cite{BLM24}, to study, in a mathematically rigorous manner, radially symmetric two-dimensional contact interactions supported on a circumference of radius $R$. From the radial symmetry it is possible to reduce the problem to one dimension (the radial one), and apply in an almost straightforward way the distributional approach of \cite{BLM24} to a contact potential at $r=R$. In the actual one-dimensional case, the usual four parameters of the point interactions can be rewritten in terms of four physical potentials strengths, corresponding to a scalar, a pseudoscalar and the two componentes of a (1+1) Lorentz electromagnetic point potential \cite{BLM24}. As it will be seen here, in the two-dimensional case with radial symmetry and a singular potential concentrated on a circumference, those four parameters will correspond to the physical strengths of a scalar and the three components of a (2+1) electromagnetic Lorentz singular potential. After obtaining the general solutions for the Dirac equation with the most general contact interaction concentrated on the circumference, we will consider some special cases, namely a purely scalar potential,    a purely electrostatic potential, a purely magnetic potential and two mixtures of a scalar and an electrostatic potential. In each of these cases we will systematically investigate the confining properties of the contact potential, understood broadly as the structure of the bound states and resonances, and the conditions under which the circular barrier becomes impenetrable.

The structure of the paper is as follows. In Section \ref{review} we review the covariant form of the Dirac equation in (2+1) dimensions with the most general external  regular potential with radial symmetry. 
In Section \ref{contact} we consider the (2+1) Dirac equation with the most general, time-independent, contact potential supported on a circumference of radius $R$. In Section \ref{eigen} we determine the equations for the energy eigenstates of the previously derived Dirac radial equation, including bound states, scattering states, and resonances. These quantities are subsequently investigated for some special cases of physical interest in Section \ref{special}. The paper ends with a discussion and conclusions in Section \ref{epilogue}


\section{Planar Dirac equation with a radial regular potential}
\label{review}

In this section we briefly review the covariant form of the Dirac equation in (2+1) dimensions with the most general external \emph{regular} potential\footnote{By this we mean a potential that is described, in the laboratory frame, by a function $f(x,t)$ that is locally integrable with respect to the space coordinates, i.e.,  $\int_K |f(x,t)|d^2 x < \infty$ for every \emph{closed} finite $K\subset \mathbb{R}^2$.}, which is formed by a scalar $B(x)$ and a vector $A_\mu(x)$,  with $x=\left(x^0,x^1,x^2\right)$ and $\mu=0,1,2$ (we are using natural units, $\hbar=c=1$):
\begin{equation}\label{regDE}
\left(i\gamma^\mu \partial_\mu -m\right)\Psi\left(x\right)={\cal I}(x)\Psi(x) ,
\quad
\text{where}
\quad
{\cal I}(x)=B(x)\mathbb{1} + A_\mu(x)\gamma^\mu\, .
\end{equation}
For convenience, we will use the following representation of the  Dirac matrices, in terms of the Pauli matrices:
\begin{equation}\label{gammapauli}
\gamma^0=\sigma_3\, , \quad  
\gamma^1=\sigma_3  \sigma_1\, \quad \text{and}\quad \gamma^2=\sigma_3  \sigma_2\, .
\end{equation}
We are interested in the case where in a particular reference frame (e.g. the laboratory frame) the potentials are independent of time, and the time-independent Dirac equation reads as follows:
\begin{equation}\label{Dirlab}
\left(\gamma^0 E + i\gamma^k \partial_k -m\right)\psi\left(\mathbf{r}\right)={\cal I}\left(\mathbf{r}\right)\psi\left(\mathbf{r}\right)\, ,\qquad k=1,2,
\end{equation}
where $\mathbf{r}=(x,y)$ is the position vector of the Dirac particle in the cartesian coordinates of the laboratory frame, $\psi\left(\mathbf{r}\right)$ is the corresponding time independent Dirac spinor and
\begin{equation}\label{genV}
{\cal I}\left(\mathbf{r}\right)=B\left(\mathbf{r}\right)\,\mathbb{1}+A_\mu\left(\mathbf{r}\right)\,\gamma^\mu\,,\qquad \mu=0,1,2.
\end{equation}
Even if the \emph{potential} in the Dirac equation (\ref{Dirlab}) is $V(\mathbf{r})=\gamma^0\,{\cal I}\left(\mathbf{r}\right)$,  we will sometimes abuse the language and will refer to ${\cal I}\left(\mathbf{r}\right)$ as the potential.

Since we are only interested in the case where the potentials have radial symmetry (in the laboratory reference frame), it is natural to use polar coordinates from now on.
The space cartesian components $A_k\left(\mathbf{r}\right)$ ($k=1,2$) of the Lorentz vector potential transform to the polar components $A_r(r,\theta)$ and $A_\theta(r,\theta)$, with $r=|\mathbf{r}|$ and $\theta$ the polar angle, as:
\begin{equation}
A_r(r,\theta)=A_1(x,y)\cos\theta +A_2(x,y)\sin\theta,\quad \label{poltransf} 
A_\theta(r,\theta)=-A_1(x,y)\sin\theta +A_2(x,y)\cos\theta\, ,
\end{equation}
where $A_1(x,y)=A_1(r\cos \theta,r\sin\theta)$, $A_2(x,y)=A_2(r\cos \theta,r\sin\theta)$, and, with a slight abuse of notation,  
$B(r,\theta)\equiv B(r\cos\theta,r\sin\theta)$, $A_0(r,\theta)\equiv A_0(r\cos\theta,r\sin\theta)$.
By defining
\begin{equation} \label{gammapol}
\gamma^r=\gamma^1\cos\theta +\gamma^2\sin\theta=
\left(\!\!
\begin{array}{cc}
0&e^{-i\theta}\\
-e^{i\theta}&0
\end{array}
\!\!\right),\quad  
\gamma^\theta=-\gamma^1\sin\theta +\gamma^2\cos\theta=
\left(\!\!
\begin{array}{cc}
0&-ie^{-i\theta}\\
-ie^{i\theta}&0
\end{array}
\!\!\right) ,
\end{equation}
we can write (\ref{genV}) in polar coordinates as 
\begin{equation}\label{Vpol}
{\cal I}(r\cos\theta,r\sin\theta)= B(r,\theta) \mathbb{1}+A_0(r,\theta)\gamma^0+A_r(r,\theta) \gamma^r + A_\theta (r,\theta) \gamma^\theta \equiv {\cal I}(r,\theta) \, .
\end{equation}
The whole potential will have \emph{radial symmetry} if all the above four potentials $B(r,\theta)$, $A_0(r,\theta)$, $A_r(r,\theta)$ and $A_\theta(r,\theta)$ depend only on $r$, and in this case we will denote them simply as $B(r), A_0(r), A_r(r)$ and $A_\theta(r)$.\footnote{In (2+1) dimensions, the magnetic field ${\cal{B}}$ is a scalar quantity (under rotations), given by the $z-component$ of the curl of a space vector lying in the $r\theta$-plane, 
$
{\cal{B}}=\left(\nabla\times \mathbf{A}\right)_z=\frac{1}{r}\left[\partial_r\left(r A_\theta\right)-\partial_\theta A_r\right]\, .
$
With radial symmetry, $\partial_\theta A_r=0$, and we have
$
{\cal{B}}=\left(\partial_r +\frac{1}{r}\right)A_\theta(r).
$
} Thus, in polar coordinates the Dirac equation (\ref{Dirlab}), with a radial potential (\ref{Vpol}), reads\footnote{To obtain the Dirac equation in polar coordinates we used (\ref{gammapol}) and the transformations of the derivatives:
\begin{equation*}
\partial_1=\left(\frac{\partial r}{\partial x^1}\right)\partial_r + \left(\frac{\partial \theta}{\partial x^1}\right)\partial_\theta=\cos\theta \partial_r - \frac{\sin\theta}{r} \partial_\theta,
\qquad
\partial_2=\left(\frac{\partial r}{\partial x^2}\right)\partial_r + \left(\frac{\partial \theta}{\partial x^2}\right)\partial_\theta=\sin\theta \partial_r +\frac{\cos\theta}{r} \partial_\theta .
\end{equation*}
}
\begin{eqnarray}\nonumber
\left(\gamma^0 E +i\gamma^r\partial_r+\frac{i}{r}\gamma^\theta \partial_\theta -m\right)\psi(r,\theta)&=&{\cal I}(r,\theta)\psi(r,\theta)\\ \label{Dirpol} 
&&\hspace{-4.5cm}=\left[B(r) \mathbb{1}+A_0(r)\gamma^0+A_r(r) \gamma^r + A_\theta (r) \gamma^\theta \right]\psi(r,\theta), \quad r>0.
\end{eqnarray}

Taking advantage of the radial symmetry, we seek for solutions of the time independent Dirac equation that are also eigensolutions of the total angular momentum operator:
\begin{equation}
J_z=L_z+\Sigma\, ,\quad {\rm{where}}\quad L_z=-i\partial_\theta\quad {\rm{and}}\quad \Sigma=\frac{1}{2}\sigma_z=\frac{1}{2}\gamma^0\, .
\end{equation} 
By writing the time independent Dirac spinor in the separated form 
\begin{equation}\label{sepDS}
\psi(r,\theta)=\beta(\theta)\Phi(r)\,,
\end{equation}
where $\beta(\theta)=\left(\begin{array}{cc} \beta_1(\theta)&0\\0&\beta_2(\theta) \end{array}\right)$ and  $\Phi(r)=\left(\begin{array}{c} \phi_1(r)\\i\phi_2(r) \end{array}\right)$, the eigenvalue equation for $J_z$ is
$J_z \left(\beta \Phi\right)=j \left(\beta \Phi\right)$, which implies
\begin{eqnarray*}
-i\partial_\theta\beta_1 +\frac{1}{2}\beta_1=j\beta_1\qquad
-i\partial_\theta\beta_2 -\frac{1}{2}\beta_2=j\beta_2
\end{eqnarray*}
and, thus,
\begin{equation}\label{eigenJ}
\beta(\theta)=
\left(\begin{array}{cc}
e^{i\left(j-\frac{1}{2}\right)\theta}&0\\
0&e^{i\left(j+\frac{1}{2}\right)\theta}
\end{array}
\right)=\left(\begin{array}{cc}
e^{il\theta}&0\\
0&e^{i\left(l+1\right)\theta}
\end{array}
\right)\,, \qquad l=j-\frac{1}{2}\in \mathbb{Z}\, .
\end{equation}
It is direct to see that $\beta$ is a unitary matrix, $\beta^\dagger \beta=\mathbb{1}$. 

Now, turning (\ref{sepDS}), with (\ref{eigenJ}), into the Dirac equation (\ref{Dirpol}), and multiplying from the left by $\beta^\dagger$, we obtain
\begin{equation}\label{Dirpol2}
\left[\left(\beta^\dagger\gamma^0\beta\right) E +i\left(\beta^\dagger\gamma^r\beta\right)\partial_r+\frac{i}{r}\left(\beta^\dagger\gamma^\theta\beta\right)\left(\beta^\dagger \partial_\theta \beta\right)-m\right]\Phi(r)=\left[\beta^\dagger {\cal I}(r,\theta)\beta\right] \Phi(r)\, .
\end{equation}
From (\ref{eigenJ}), (\ref{gammapol}) and (\ref{gammapauli}) we have 
\begin{eqnarray}\nonumber
\beta^\dagger\gamma^0\beta&=&\gamma^0, \qquad\qquad
\beta^\dagger\gamma^r\beta=\gamma^1, \qquad \qquad
\beta^\dagger\gamma^\theta\beta=\gamma^2, \\  [0.5ex]
\beta^\dagger {\cal I}(r,\theta)\beta&=& B(r)\mathbb{1}+A_0(r)\gamma^0+A_r(r)\gamma^1+A_\theta(r) \gamma^2 \equiv \tilde{\cal I}(r), 
\label{poltransf}
\\  [0.5ex]
\beta^\dagger \left(\partial_\theta \beta \right)&=&i\left(l+\frac{1}{2}\right)\mathbb{1}-\frac{i}{2}\gamma^0.
\nonumber
\end{eqnarray}
Substituting the above results into the equation (\ref{Dirpol2}), we obtain the radial Dirac equation 
\begin{equation}\label{radialdirac}
\left[\gamma^0 E  +i\gamma^1 \left(\partial_r+\frac{1}{2r}\right)-\frac{1}{r}\left(l+\frac{1}{2}\right)\gamma^2 -m\right]\Phi(r)= \tilde{\cal I}(r) \Phi(r)\,,
\end{equation}
which can be written in terms of an \emph{effective} potential:
\begin{equation}\label{Dirac1d}
\left[\gamma^0 E +i\gamma^1 \partial_r-m\right]\Phi(r)={\cal I}_{\rm{eff}}(r) \Phi(r)\, ,\qquad r>0,
\end{equation}
where
\begin{eqnarray}\nonumber
{\cal I}_{\rm{eff}}(r)&=& \tilde{\cal I}(r) -\frac{i}{2r}\,\gamma^1+\frac{1}{r}\left(l+\frac{1}{2}\right)\,\gamma^2\\
\label{Veff}
&=& B(r)\mathbb{1}+A_0(r)\gamma^0+A_r(r)\gamma^1+A_\theta(r) \gamma^2 -\frac{i}{2r}\,\gamma^1+\frac{1}{r}\left(l+\frac{1}{2}\right)\,\gamma^2\, .
\end{eqnarray}


\subsection{Conservation of the Dirac current}


The Dirac current in (2+1) dimensions, in cartesian components, is given by
\begin{equation}\label{current}
j^\mu(x)=\Psi^\dagger(x)\gamma^0 \gamma^\mu \Psi(x), \qquad x=\left(x^\mu\right),\quad \mu=0,1,2\,.
\end{equation}
The current conservation states that
$\partial_\mu j^\mu(x)=0$, 
which, in the cartesian coordinates of the laboratory frame, reads as
\begin{equation}\label{divj}
\partial_k j^k (x^\mu )=-\partial_0 j^0 (x^\mu )\, .
\end{equation}
Integrating the above on a arbitrary circle $S_\rho$ of radius $\rho$, and using the divergence theorem, we obtain (below, $\mathbf{\hat{r}}$ is the unit vector in the radial direction):
\begin{equation} 
\int_{S_\rho} \mathbf{\nabla}\cdot \mathbf{j}\left(x^\mu\right) ds=\oint_{C_\rho} \mathbf{j}\left(x^\mu\right)\cdot \hat{\mathbf{r}} \,dl = -\frac{d}{d t}\int_{S_\rho} j^0\left(x^\mu\right) ds \label{divj}
 2\pi \rho \,j^r(\rho,t)=  -\frac{d}{d t} Q_\rho (t)\, ,
\end{equation}
where $Q_\rho(t)$ is the total probability to find the Dirac particle enclosed inside the circumference  $C_\rho$ of radius $\rho$. Above, we identified the radial component of the Dirac current, $j^r\left(\rho,t\right)$, and used the fact that the radial symmetry implies that it does not depend on $\theta$ (see below). 

Now we restrict ourselves to the stationary case. From the radial symmetry, we have
\begin{equation} \label{statcurr}
j^\mu = \Phi^\dagger \beta^\dagger \gamma^0\gamma^\mu \beta \Phi
=\Phi^\dagger \left(\beta^\dagger \gamma^0\beta\right) \left(\beta^\dagger\gamma^\mu \beta\right) \Phi
= \Phi^\dagger \gamma^0 \left(\beta^\dagger\gamma^\mu \beta\right) \Phi, \quad \mu=0,1,2.
\end{equation}
The explicit form of the  component $j^0$ is
\begin{equation*}
j^0=\Phi^\dagger \Phi = \overline{\Phi}\gamma^0 \Phi\, , \qquad \overline{\Phi}\equiv\Phi^\dagger \gamma^0\, ,
\end{equation*}
and, for the cartesian space components, we have
\begin{eqnarray*}
j^1=\overline{\Phi} \left(\beta^\dagger \gamma^1\beta\right) \Phi &=&\overline{\Phi} \left(\gamma^1\cos \theta-\gamma^2 \sin\theta\right)\Phi = j^r\,\cos\theta - j^\theta\,\sin\theta,
\\ [0.5ex]
j^2=\overline{\Phi} \left(\beta^\dagger \gamma^2\beta\right) \Phi &=&\overline{\Phi} \left(\gamma^1\sin \theta+\gamma^2 \cos\theta\right)\Phi = j^r\,\sin\theta + j^\theta \cos\theta\, ,
\end{eqnarray*}
where, from (\ref{gammapol}) and (\ref{statcurr}), we have identified the $r$ and $\theta$-components of the stationary Dirac current as
\begin{eqnarray} \label{jr}
j^r&=& j^1 \cos\theta + j^2\sin\theta=\overline{\psi} \gamma^r \psi =\overline{\Phi}\gamma^1 \Phi, 
\\  [0.5ex]
j^\theta&=&-j^1\sin\theta + j^2\cos\theta 
= \overline{\psi} \gamma^\theta \psi =\overline{\Phi}\gamma^2 \Phi  .
\label{jtheta}
\end{eqnarray}
From (\ref{jr}) and (\ref{jtheta}) it is clear that both the current's polar components $j^r$ and $j^\theta$ do not depend on the angular coordinate $\theta$, as mentioned earlier. Thus, we write $j^r(r,\theta)\equiv j(r)$ and $j^\theta(r,\theta)\equiv j^\theta(r)$.
From the above we have that along the angular direction  both the \emph{stationary} currents $j^r$ and $j^\theta$ are constant and, from (\ref{divj}),  in the stationary case we have that the quantity $r j^r$ vanishes, which implies that
$j^r(r) = 0$, $\forall\, r>0$.
In particular, for any radius $r=\rho>0$:
\begin{equation}\label{cc}
j^r (\rho^- )=j^r (\rho^+ )=j^r(\rho)=0, \qquad {\rm{where}}\quad j^r (\rho^\pm )\equiv \lim_{r\to \rho^\pm} j^r(r)\, .
\end{equation}

Summarizing the results obtained so far, we have that, for an external, \emph{regular}, and radial symmetric potential, ${\cal I}(r,\theta)=\left[B(r)\mathbb{1}+A_0(r)\gamma^0+A_r(r) \gamma^r+A_\theta(r)\gamma^\theta\right]$, the radial spinor $\Phi(r)$ must solve the one-dimensional stationary Dirac equation (\ref{Dirac1d}) with the effective one-dimensional potential (\ref{Veff}). 
Conservation of the Dirac current in the original two-dimensional problem implies that the one-dimensional (radial) stationary current $j^r(r)$, given by (\ref{jr}), is \emph{continuous} (and vanishes) at any $r$, and therefore satisfies condition (\ref{cc}).

Equation (\ref{Dirac1d}) is \emph{mathematically analogous} to the Dirac equation in (1+1) dimensions, with an external potential ${\cal I}_{\rm{eff}}(r)$ (see equation (3) at p. 4 of \cite{BLM24}). However, 
(\ref{Dirac1d}) holds only for $r>0$. In the next Section we will see how to extend it as a distributional equation on the interval $r\in [0,\infty)$, and which condition the Dirac spinor $\psi$ must satisfy at $r=0$. The analogous roles of the potentials $A_1(x)$ and $W(x)$ of the \emph{regular, true one-dimensional case} (see the equation (2) of \cite{BLM24}) will be played in (\ref{Dirac1d}) by the following combinations
\begin{eqnarray}\label{anA1}
A_1 (x^1 ) \longrightarrow A_r(r) -\frac{i}{2r},
\qquad
W (x^1 ) \longrightarrow -A_\theta(r) -\frac{1}{r}\left(l+\frac{1}{2}\right) .
\end{eqnarray}
It is worth to observe that in (\ref{Dirac1d}) the \emph{effective} one-dimensional potential in the radial direction, $\gamma^0 {\cal I}_{\rm{eff}}(r)$, is \emph{not} Hermitian, due to the presence of the imaginary quantity $-\frac{i}{2r}$ added to the potential $A_r(r)$ in (\ref{anA1}). However, this is immaterial, since for the distributional method of \cite{BLM24} to be applicable in the present case it suffices that the current be \emph{continuous} at $r=R$. Finally, it is worth to note that the original \emph{potential} $\gamma^0 {\cal I}(r,\theta)$ is in fact Hermitian, and the Lorentz vector $j^\mu$ is conserved.
 

\section{Dirac equation with radial contact potentials at $r=R$}
\label{contact}

Now we consider the Dirac equation with the most general, time-independent, contact potential supported on a circumference of radius $R$ in the laboratory frame. Since now the potential is singular, the products between the potentials and the spinors $\Phi(r)$ appearing in (\ref{Dirac1d}) are not well defined. Guided by the analogy to the one-dimensional case, we follow closely the steps of reference \cite{BLM24} and write the time independent Dirac equation as
\begin{equation}\label{Dirac1dsing}
\left[\gamma^0 E +i\gamma^1 \partial_r-m\right]\Phi(r)=D[\Phi](r)+
{\cal I}^{\rm{reg}}_{\rm{eff}}(r) \Phi(r)\, ,\qquad r>0.
\end{equation}
where
\begin{equation}\label{Veffreg}
{\cal I}^{\rm{reg}}_{\rm{eff}}(r)=-\frac{i}{2r}\gamma^1+\frac{1}{r}\left(l+\frac{1}{2}\right) \gamma^2\, 
\end{equation}
is the regular part of the effective potential. Since this term defines a regular distribution on $[0,\infty)$, which includes the origin, the equation (\ref{Dirac1dsing}), understood as a distributional equation, can be extended to include the origin, which will be assumed from now on.\footnote{\label{footreg} In polar coordinates, a radial function $f(r), r>0$ is said to define a \emph{regular} distribution if it is locally integrable on $(0,\infty)$ , i.e., if $\int_K  |f(r)|\,r dr <\infty$ for any \emph{closed} finite interval $K\subset (0,\infty)$, in which case we denote $f\in {\cal L}^1_{\text{loc}}\left(0,\infty \right)$. Due to the presence of the Jacobian factor $r$  in the integral, the potential (\ref{Veffreg}) is locally integrable also on $\left[0,\infty \right)$.}  The first term in the right hand side of (\ref{Dirac1dsing}) is the \emph{contact} term, and is a $2\times 1$ matrix, yet unknown, whose elements are \emph{singular} (not regular) distributions, related to the singular potentials concentrated at $r=R$. 
The distribution space we consider here is the Schwartz space ${\cal S}^\prime$, which consists of continuous linear functionals acting on the Schwartz space ${\cal S}$ of infinitely differentiable strongly decreasing test functions. 
For details, we refer the reader to reference \cite{BLM24}. The distribution $D[\Phi](x)$ will be determined from the following three basic requirements (adapted from \cite{BLM24}):

\begin{itemize}

\item[\textbf{(R1)}] $D[\Phi](r)$ must be a distribution \emph{concentrated} at $r=R$, i.e., it must equals the zero distribution in $\mathbb{R}_+\backslash \{R\}$.

\item[\textbf{(R2)}] The two components of the radial Dirac spinor $\Phi(r)$ must correspond to \emph{regular distributions} on $\mathbb{R}_+=[0,\infty)$, i.e., they must be slow growth functions and locally integrable (in the Lebesgue sense) on $\mathbb{R}_+$. This is equivalent  to saying that $\Phi(r)$ must have \emph{order} $s_\Phi\leq -1$ on any closed interval $K\ni R$.

\item[\textbf{(R3)}] The Dirac current must be conserved across $r=R$. In the stationary case this implies that $j^r\left(R^-\right)=j^r\left(R^+\right)=0$. 

\end{itemize} 

The definition we are using above for the \emph{order} of a distribution on a closed, finite interval $K\subset \mathbb{R_+}$, is the same of \cite{BLM24}, adapted to the semi-infinite axis $\mathbb{R_+}$. Briefly, with $R\in K$, the order of the Dirac delta distribution $\delta(r-R)$ on $K$ is $s=0$. Taking the derivative of any distribution increases its order on $K$ by $+1$, whereas  taking an indefinite integral (a primitive) decreases its order by $-1$. The  order of a sum of distributions on $K$ is the largest order of its terms. A distribution has order $s=-\infty$ on $K$ if it is infinitely differentiable on $K$. A distribution has order $s=-2$ on $K$ if on this interval it coincides with a continuous but not differentiable (in the ordinary sense) function. A distribution of order $s=-1$ corresponds to a discontinuous function, and may or may not be regular. A distribution having order $s\geq 0$ on $K$ is always singular. For more details, we refer the reader to \cite{BLM24}.

From requirement \textbf{(R2)}, and balancing the orders on both sides of (\ref{Dirac1dsing}), the  order of the interaction distribution $D[\Phi]$ on any interval $K\ni R$ must be $s_{D[\Phi]}\leq 0$. This same  requirement also sets a maximum for the order of the spinor $\Phi(r)$ on a closed finite interval $I_0$ containing the origin, but not containing $R$, i.e., $I_0=[0,b]$, $b<R$. Since on $I_0$ the contact distribution $D[\Phi]$ equals the zero distribution, and the term (\ref{Veffreg}) has order $\leq -1$ (since it is regular on $I_0$, see footnote \ref{footreg}), by balancing the orders on both sides of   (\ref{Dirac1dsing}) we conclude that  $s_{\Phi^\prime}\leq -1$ on $I_0$ and, consequently, $s_\Phi\leq -2$ on $I_0$. This means that the spinor $\Phi$ \emph{must be continuous} on $I_0$ and, therefore, it \emph{must be bounded at the origin}.\footnote{By a \emph{continuous} distribution on a closed, finite interval $ K\subset [0,\infty)$, we are referring to a regular distribution which coincides with a continuous ordinary function almost everywhere on $K$.} This is another way to justify, in a mathematically rigorous way, the usual assumption of boundedness of the Dirac spinor at $r=0$.

Theorem 3.5-2 of \cite{Zem87} states that a distribution that is concentrated at a single point $r=R$, and that has singular order $s$ on a closed finite interval $K\ni R$, must in this interval coincide with a finite linear combination of Dirac deltas and its derivatives, up to order $s$. Since on $K$ the order of $\delta(r-R)$ is zero, we must have $s_{D[\Phi]}\geq 0$. We have that, on $K$, $s_\Phi\leq -1$ (from \textbf{(R2)}) and, consequently,  $s_{\Phi^\prime}\leq 0$. Now, by balancing the orders on both sides of (\ref{Dirac1dsing}), we have that $s_{D[\Phi]}=s_{\Phi^\prime}\leq 0$, and therefore $s_{D[\Phi]}= 0$ on $K$. Requirements \textbf{(R1)} and \textbf{(R2)} thus imply that  
\begin{equation}\label{Ddist}
D[\Phi](r)=\Omega[\Phi]\, \delta(r-R), 
\end{equation} 
where the $2\times 1$ matrix $\Omega[\Phi]$ does not depend on the coordinate $r$, and will have a functional (linear) dependence of the spinor $\Phi$. The form of $\Omega[\Phi]$ will completely specify the contact term, and below it will be found from requirement \textbf{(R3)}. As  will be shown, the obtained form of $\Omega[\Phi]$ will imply matching conditions that $\Phi(r)$ must satisfy at $r=R$, similar to the true one-dimensional case. Also, it will provide the contact term $\Omega[\Phi]\, \delta(r-R)$ explicitly as a well defined distribution defined on the semi-axis $\mathbb{R}_+$. 

To determine $\Omega[\Phi]$ we proceed exactly in the same way as in \cite{BLM24}. By taking a distributional indefinite integral of equation (\ref{Dirac1dsing}) on $K\subset \mathbb{R}_+$, $K\ni R$, we obtain
\begin{equation}\label{primi}
\left[\gamma^0 E-m\right]\Phi^{(-1)}(r) +i\gamma^1 \Phi(r)=\Omega[\Phi]\theta(r-R)+
F(r) + c\, ,
\end{equation}
where $\Phi^{(-1)}$ is any primitive of $\Phi$, $F(r)$ is \emph{a primitive} of the product ${\cal I}^{\rm{reg}}_{\rm{eff}}(r)\Phi(r)\in {\cal L}^1_{\rm{loc}}(K)$ and $c$ is a constant and arbitrary column matrix. Similarly to the reasoning in \cite{BLM24}, we conclude that both functions, $\Phi^{(-1)}$  and $F(r)$, are \emph{continuous} on $K$.\footnote{That ${\cal I}^{\rm{reg}}_{\rm{eff}}(r)\Phi(r)$ is regular on $K\ni R$ is clear from the fact that $\Phi$ is regular (from assumption \textbf{R2}) and ${\cal I}^{\rm{reg}}_{\rm{eff}}(r)$ is continuous at $r=R$. Thus, this product is a regular distribution and, therefore, any of its primitives is a continuous function. By the way, ${\cal I}^{\rm{reg}}_{\rm{eff}}(r)\Phi(r)$ is also regular on $[0,b], b<R$, since on this interval ${\cal I}^{\rm{reg}}_{\rm{eff}}(r)$ is regular and $\Phi(r)$ is continuous.} By isolating the term $i\gamma^1\,\Phi(r)$ in the above equation, we find that $\Phi(r)$ must have well defined lateral limits at $r=R$. By considering the limits $r\to R^\pm$ in (\ref{primi}) and taking the difference, we obtain:
\begin{equation}\label{match}
i\gamma^1\left[\Phi\left(R^+\right)-\Phi\left(R^-\right)\right]=\Omega[\Phi]\, .
\end{equation}  
Now the functional coefficient $\Omega[\Phi]$ will be determined from requirement \textbf{(R3)}: 
\begin{equation}\label{cc2}
j^r (R^- )=j^r (R^+ ) , \quad {\rm{with}}\quad j^r(r)=\overline{\Phi}\gamma^1\Phi\, .
\end{equation}
The problem of determining $\Omega[\Phi]$ from (\ref{match}), by using the current condition (\ref{cc2}), is mathematically analogous to the one-dimensional problem treated in \cite{BLM24}. Therefore, we just collect the  results from \cite{BLM24}, replacing the analogous quantities to the present problem:
\begin{equation}\label{singterm}
\Omega[\Phi]=\left(B\mathbb{1} + A_0 \gamma^0 +A_r \gamma^1 + A_\theta \gamma^2\right)\frac{\Phi\left(R^-\right)+\Phi\left(R^+\right)}{2}\, ,
\end{equation}
where the real \emph{constants} $B, A_0, A_r$ and $A_\theta$ are identified with the \emph{strengths of the singular potentials} associated to the contact term supported on $r=R$, namely, from (\ref{Ddist})
\begin{equation}\label{singpot}
B(r)=B\,\delta(r-R),\ \ A_0(r)=A_0\, \delta(r-R),\ \  A_r(r)=A_r\,\delta(r-R)\ \  \text{and}\ \  A_\theta(r)=A_\theta \,\delta(r-R).
\end{equation}
The radial Dirac equation (\ref{Dirac1dsing}), with the contact and the regular interaction terms given explicitly, is: 
\begin{eqnarray}\nonumber
\left(\gamma^0 E +i\gamma^1 \partial_r-m\right)\Phi(r)&=&\Omega[\Phi]\delta(r-R)  + {\cal I}^{\rm{reg}}_{\rm{eff}}(r) \Phi(r)\\ \nonumber
&=&\left(B\mathbb{1} + A_0 \gamma^0 +A_r \gamma^1 + A_\theta \gamma^2\right) \delta(r-R)\, \frac{\Phi\left(R^-\right)+\Phi\left(R^+\right)}{2}\\ 
&& +\left[-\frac{i}{2r}\gamma^1+\frac{1}{r}\left(l+\frac{1}{2}\right) \gamma^2\right]\Phi(r) \, .
\label{Dirac1dsingfin}
\end{eqnarray}
A \emph{permeable} contact interaction is characterised by the condition (see equation (55) of~\cite{BLM24})
\begin{equation}\label{permcond}
A_r\neq 0\quad\text{or}\quad B^2-4-A_0^2+A_\theta^2\neq 0\, ,
\end{equation}
which means that the solutions at $r=R^-$ and $r=R^+$ are connected by the following matching conditions 
\begin{equation}\label{Lambda}
\Phi\left(R^+\right)=\Lambda \Phi\left(R^-\right),\quad {\rm{with}}\quad \Lambda=e^{i\varphi}
\left(
\begin{array}{cc}
a&ib\\
-ic&d
\end{array}
\right),\quad ad-bc=1\, , \quad \varphi\in [0,\pi)\,,
\end{equation}
where the $\Lambda$ matrix is written in terms of the physical strengths as\footnote{Observe that, when collecting the results from equation (57) of \cite{BLM24}, we should replace  $W \to -A_\theta$.}
\begin{equation}\label{lambdafields}
\Lambda =\tfrac{e^{i \arg\left(B^2-4-A_0^2+A_r^2+A_\theta^2+4 i A_r\right)}}{\sqrt{\left(B^2-4-A_0^2+A_r^2+A_\theta^2\right)^2+16A_r^2} }  \left(
\begin{array}{cc}
 { \scriptstyle -B^2+A_0^2-A_r^2
   -(A_\theta+2)^2 }& { \scriptstyle 4i (A_0-B) } \\
 { \scriptstyle 4i (A_0+B) } &
   { \scriptstyle -B^2+A_0^2-A_r^2
   -(A_\theta-2)^2 } \\
\end{array}
\right)\, .
\end{equation}
Yet, to have a unique set of $\Lambda$-parameters $\varphi, a,b,c,d$, given a set of parameters $B, A_0,$ and $A_r$ and $A_\theta$ satisfying the permeability conditions (\ref{permcond}), we can write the relationships:
\begin{eqnarray}
\label{fp}
\varphi&=&\tan^{-1}\left(\tfrac{4 A_r}{B^2-4-A_0^2+A_r^2+A_\theta^2}\right) ,\quad \varphi\in [0,\pi)\, ,\\ [0.6ex] 
\label{ap}
a&=& \pm \tfrac{A_0^2-A_r^2-B^2
   -(A_\theta+2)^2}{\sqrt{\left(B^2-4-A_0^2+A_r^2+A_\theta^2\right)^2+16A_r^2}}, \qquad   
b= \pm \tfrac{4 (A_0-B)}{\sqrt{\left(B^2-4-A_0^2+A_r^2+A_\theta^2\right)^2+16A_r^2}}, 
\\  [0.6ex] 
c&=& \pm \tfrac{-4 (A_0+B)}{\sqrt{\left(B^2-4-A_0^2+A_r^2+A_\theta^2\right)^2+16A_r^2}}, \qquad  
\label{dp}
d= \pm \tfrac{A_0^2-A_r^2-B^2
   -(A_\theta-2)^2}{\sqrt{\left(B^2-4-A_0^2+A_r^2+A_\theta^2\right)^2+16A_r^2}} ,
\end{eqnarray}
where in the above expressions the plus (minus) sign must be taken if $A_r> 0$ ($A_r<0$); if $A_r=0$ we must take the same sign of $B^2-4-A_0^2+A_\theta^2$ (see \cite{BLM24}). 

When the strength potentials are such that the permeability condition (\ref{permcond}) is not met, the circular wall at $r=R$ is impenetrable, and some (or all) of the $\Lambda$ matrix parameters becomes infinite. In this case, the inner and  outer solutions are independent, the (anti)particle is completely confined in the inner (outer) region of the circumference of radius $R$, and it cannot be transmitted to the outer (inner) region. 
The set of allowed energies for the (anti)particle confined in the inner region will depend on the boundary conditions at $r=R$, which in turn depend on the values of the strength potentials. In the outer region all  energies $|E|>m$ are allowed; the  $E=\pm m$ cases may or may not be allowed in the outer region, as we will see later.
In the following sections, we will use the matching conditions (\ref{Lambda}), together with the form  (\ref{lambdafields}) of the $\Lambda$ matrix in terms of the physical strengths, to study bound, scattering, and resonant states for various choices of the contact potentials. Cases that give rise to an impermeable wall can be obtained by considering the corresponding limits on the $\Lambda$ parameters.


\section{Energy eigenstates}
\label{eigen}

In this section we will determine the energy eigenstates for the radial Dirac equation (\ref{Dirac1d}), for a given angular momentum $l>0$. Consider the spinorial form of the solutions, for the inner ($r<R$, $\alpha=i$) or the outer ($r>R$, $\alpha=o$) regions
\begin{equation}\label{estado}
\Phi_{\alpha}(r) = \left(
\begin{array}{c}
\phi_{1,\alpha}(r)
\\ 
\displaystyle i \phi_{2,\alpha}(r)
\end{array}\right) .
\end{equation}
The spinors $\Phi_{\alpha}(r)$ are solutions of equation (\ref{radialdirac}), with $\tilde{\cal I}(r)=0$ (the \emph{free} equation). In terms of the spinor components  $\phi_{k,\alpha}(r)$, $k=1,2$, $\alpha=i,o$, that equation reads
\begin{equation}\label{freecoupled}
- \frac{d\phi_{1,\alpha} }{dr} + \frac{l}{r}\ \phi_{1,\alpha}  =  (E+m) \  \phi_{2,\alpha} ,
\quad 
\frac{d\phi_{2,\alpha}}{dr} +\frac{l+1}{r}\ \phi_{2,\alpha} = (E-m) \ \phi_{1,\alpha} ,\quad \forall \,r \neq R.
\end{equation}
When $|E|\neq m$, by isolating $\phi_{2,\alpha}$ from the first of equations (\ref{freecoupled}) and substituting it into the second equation, we obtain a Schrödinger-like equation for $\phi_{1,\alpha}$, valid  $\forall \,r \neq R$,
\begin{eqnarray}\nonumber
   -\frac{d^2\phi_{1,\alpha} }{dr^2}-\frac{1}{r}\frac{d\phi_{1,\alpha} }{dr}+\frac{l^2}{r^2}\phi_{1,\alpha} &=&\left(E^2-m^2 \right)\phi_{1,\alpha},
   \\    \label{freedecoupled}
r^2\,\frac{d^2\phi_{1,\alpha} }{dr^2}+r\,\frac{d\phi_{1,\alpha} }{dr}+\left(p^2 r^2-l^2\right)\phi_{1,\alpha}   &=& 0\, , 
\end{eqnarray}
where we defined $p=\sqrt{E^2-m^2}$. The solution of (\ref{freedecoupled}) is (see equation (2.13) of \cite{KNN22})\footnote{Note that here $p$ is the same for both the inner and the outer regions.}
\begin{eqnarray} 
\phi_{1,\alpha} (r)&=&a_\alpha \,J_l \left(p\,r \right)+b_\alpha\, Y_l \left(p\,r \right) =\tilde{a}_\alpha \,H_l^{(1)} \left(p\,r \right)+\tilde{b}_\alpha\, H_l^{(2)} \left(p\,r \right),\quad
\alpha=i,o\, ,
\label{phi1} 
\end{eqnarray}
where $J_l$($H_ l^{(1)}$) and $Y_l$($H_ l^{(2)}$) are, respectively, the Bessel (Hankel) functions of the first and of the second kind. Substituting this solution into the first of equations (\ref{freecoupled}), we obtain $\phi_{2,\alpha} (r)$ (see also \cite{KNN22}):
\begin{equation}
\phi_{2,\alpha} (r)=\frac{p}{E+m}\left[a_\alpha J_{l+1} (pr )+b_\alpha Y_{l+1} (pr )\right]
=\frac{p}{E+m}\left[\tilde{a}_\alpha H_{l+1}^{(1)} (pr)+\tilde{b}_\alpha H_{l+1}^{(2)} (pr )\right], \quad
\alpha=i,o .\label{phi2} 
\end{equation}
Thus, the radial part of the energy and angular momentum eingenstates (\ref{estado}) of the free Dirac particle in the inner and outer regions can be written in the form 
\begin{eqnarray}\nonumber
\Phi_{\alpha}&=& a_\alpha 
\left(
\begin{array}{c}
J_l(p\, r)\\
\frac{i p}{E+m} \,J_{l+1}(p\, r)
\end{array}
\right)
+
b_\alpha 
\left(
\begin{array}{c}
Y_l(p\, r)\\
\frac{i p}{E+m} \,Y_{l+1}(p\, r)
\end{array}
\right)\\ [1ex]
\label{inout}
&=&\tilde{a}_\alpha 
\left(
\begin{array}{c}
H^{(1)}_l(p\, r)\\
\frac{i p}{E+m} \,H^{(1)}_{l+1}(p\, r)
\end{array}
\right)
+
\tilde{b}_\alpha 
\left(
\begin{array}{c}
H^{(2)}_l(p\, r)\\
\frac{i p}{E+m} \,H^{(2)}_{l+1}(p\, r)
\end{array}
\right), \qquad \alpha=i,o\, . 
\end{eqnarray}
Since the inner solution $\Phi_{i}(r)$ must be bounded at the origin (see the discussion regarding requirement \textbf{(R2)} in Section \ref{eigen}), and this holds only for the spinor whose elements are Bessel functions of the first kind, from now on we will consider only the  first line of equations (\ref{inout}), with $b_i=0$, as the solution for the \emph{inner region}.     


\subsection{Critical and supercritical states ($|E|=m$)}


The states for which $E=m$ or $E=-m$ are called, respectively, \emph{critical} and \emph{supercritical} states and are important since, as we will see later, by varying the strengths of the potential a resonance can  transform into a bound state, and vice versa. In the former case we say that a bound state is captured (or absorbed) from the continuum, and in the latter, that a bound state was emitted into the continuum. Captures or emissions of bound states from and to the continuum always occur at a critical or supercritical energy.  To solve the Dirac equation for these states we must consider the equation (\ref{freecoupled}) with $E=\pm m$.

\subsubsection{Critical states}
\label{criticalstates}
The critical states ($E=m$) solve the equation (\ref{freedecoupled}), with $p=0$ (an Euler equation),
\begin{equation}\label{decrit}
r^2\,\frac{d^2\phi_{1,\alpha} }{dr^2}+r\,\frac{d\phi_{1,\alpha} }{dr}-l^2\phi_{1,\alpha}   = 0\, , \qquad \forall \,r \neq R\, ,
\end{equation}
and the first of equations (\ref{freecoupled}), with $E=m$:
\begin{equation}\label{f2crit}
- \frac{d\phi_{1,\alpha} }{dr} + \frac{l}{r}\, \phi_{1,\alpha}  =  2m  \, \phi_{2,\alpha}\, .\end{equation}
The solutions of (\ref{decrit}) have the form $r^{\pm l}$, but since we only consider $l\geq 0$ and require the Dirac spinor be bounded at $r=0$, only the solution $r^l$  is admissible in the inner region. We also require the Dirac spinor to be bounded at infinity (a physical requirement), so the only  admissible solution for the outer region is $r^{-l}$. Therefore, the solutions for (\ref{decrit}) in the inner and outer regions are
\begin{equation}\label{critio}
\phi_{1,i}=c_i\, r^l\quad\text{and}\quad \phi_{1,o}=c_o\, r^{-l}\, 
\end{equation}
with $c_i$ and $c_o$ being arbitrary constants. substituting these solutions into (\ref{f2crit}), we obtain $\phi_{2,i}=0$ and $\phi_{2,o}= c_o\,\frac{l}{m}\, r^{-(l+1)}$. Therefore, the solutions  for critical states in the inner and outer regions are 
\begin{equation}\label{critsol}
\Phi_i^\text{crit} = c_i 
\left(
\begin{array}{c}
r^l\\
0
\end{array}
\right)\quad\text{and}\quad 
\Phi_o^\text{crit}=c_o\left(
\begin{array}{c}
r^{-l}\\
\frac{i\,l}{m}r^{-(l+1)}
\end{array}
\right)\, .
\end{equation}
These two solutions must match at $r=R$ using the matrix $\Lambda$, according to (\ref{Lambda}),
\begin{equation}
\Phi^\text{crit}_o(R)=\Lambda\, \Phi^\text{crit}_i (R)\, ,
\end{equation}
that results in
\begin{equation}
c_o
\left(
\begin{array}{c}
R^{-l}\\
\frac{i\,l}{m}\, R^{-(l+1)}
\end{array}
\right)=c_i\,\Lambda 
\left(
\begin{array}{c}
R^{l}\\
0
\end{array}
\right)\, .
\end{equation}
This equation has non-trivial solutions for $c_{i,o}$ if, and only if, the following secular equation is satisfied by the parameters of the matrix $\Lambda$ 
\begin{equation}\label{critcond}
c+a\,\frac{l}{m\,R}=0\,.
\end{equation}
From (\ref{critsol}) we observe that critical states with $l=0,1$ are not true bound states: they are \emph{quasi-bound} states, since they are bounded or vanish at infinity, but are not square integrable.\footnote{For integrability and square integrability, the integration measure  $r dr$ must be taken into account here.} For $l\geq 2$ the critical states are true bound states.

\subsubsection{Supercritical states}
\label{supercritical}
To obtain the supercritical states ($E=-m$) we isolate $\phi_{1,\alpha}$ in the second of the equations (\ref{freecoupled}) and substitute it in the first, obtaining another  Euler equation
\begin{equation}
r^2 \frac{d^2}{dr^2}\phi_{2,\alpha} + r \frac{d}{dr}\phi_{2,\alpha} -\left(l+1\right)^2\phi_{2,\alpha}=0\, ,
\end{equation} 
whose solutions are of the form $r^{\pm(l+1)}$. Similarly to the previous case, for the inner region the solution with the positive sign is the only admissible one, while for the outer region we need only consider the solution with the negative sign. The corresponding solutions for $\phi_{1,\alpha}$ are obtained from the second of the equations (\ref{freecoupled}). Therefore, the admissible supercritical solutions in the inner and outer regions are
\begin{equation}\label{supersol}
\Phi_i^\text{super} = c_i 
\left(
\begin{array}{c}
-\frac{(l+1)}{m}\,r^l\\
i\,r^{l+1}
\end{array}
\right)\quad\text{and}\quad 
\Phi_o^\text{super}=c_o\left(
\begin{array}{c}
0\\
i\,r^{-(l+1)}
\end{array}
\right)\, .
\end{equation}
Again, to match  solutions at $r=R$ we use (\ref{Lambda}),  $\Phi^\text{super}_o(R)=\Lambda\, \Phi^\text{super}_i(R)$, and we obtain the following secular equation for the existence of non-trivial solutions:
\begin{equation}\label{supercond}
b+a\,\frac{(l+1)}{m R}=0\,.
\end{equation}
From (\ref{supersol}) we observe that, with the exception of $l=0$, the supercritical states correspond to true bound states. Let us emphasize that the phase factor $e^{i \varphi}$ of the matrix $\Lambda$  does not enter the conditions (\ref{critcond}) and (\ref{supercond}) for the existence of critical or supercritical states. 

Using the equations (\ref{ap})-(\ref{dp}) we can write the conditions for the existence of critical and supercritical states,  (\ref{critcond}) and (\ref{supercond}), in terms of the physical parameters $B$ and $A_\mu$ ($\mu=0,1,2$).

%
%

\subsection{Bound states ($|E|< m$)}

The bound states are characterized by $|E|<m$, and in this case $p=i q$, with $q=\sqrt{m^2-E^2}>0$. For the outer region, it is more convenient to consider the solution in terms of Hankel functions, since it is easier to explore their asymptotic behavior as $r\to\infty$. The Hankel functions behave asymptotically as
\begin{equation} 
H^{(1)}_l(iq\,r) \approx \sqrt{\frac{2}{i\pi q\,r} }\, e^{-q r -\frac{i}{4}(2 l+1) \pi}\, ,\qquad 
\label{asympH}
H^{(2)}_l(iq\,r) \approx \sqrt{\frac{2}{i\pi q\,r} }\, e^{qr -\frac{i}{4}(2 l-3) \pi}\, .
\end{equation}
From the above we see that the Hankel function $H^{(2)}_l(iq r)$ explodes as $r$ goes to infinity, while $H^{(1)}_l(i q r)$ gives an evanescent wave. So, for bound states, we must have $\tilde{b}_o=0$ in the second line of equations (\ref{inout}). The inner and outer solutions must be matched at $r=R$ by using  (\ref{Lambda}) :
\begin{eqnarray}
\tilde{a}_o 
\left(
\begin{array}{c}
H^{(1)}_l(iq R)\\
\frac{-q}{E+m} \,H^{(1)}_{l+1}(iq R)
\end{array}
\right)=a_i\,\Lambda  
\left(
\begin{array}{c}
J_l(i q R)\\
\frac{-q}{E+m} \,J_{l+1}(iq R)
\end{array}
\right)\, ,
\end{eqnarray}
that gives
\begin{equation}\label{linsys}
{\cal M}(iq R)
\left(\begin{array}{c}
\tilde{a}_o\\
a_i
\end{array}
\right)=\left(\begin{array}{c}
0\\
0
\end{array}
\right)\, ,
\end{equation}
where ${\cal M}(iq R)=\left(\,{\cal C}_o(iq R)\ {\cal C}_i(iq R)\,\right)$ is the matrix whose columns are 
\begin{equation}
{\cal C}_o(iq R) = \left(\begin{array}{c}
H^{(1)}_l(iq\, R)\\ [0.5ex]
\frac{-q}{E+m} \,H^{(1)}_{l+1}(iq R)
\end{array}
\right) \quad \text{and} \quad {\cal C}_i(iq R)= -\Lambda  
\left(
\begin{array}{c}
J_l(iq R)\\ [0.5ex]
\frac{-q}{E+m} \,J_{l+1}(iq\, R)
\end{array}
\right) \, .
\end{equation}
The condition for (\ref{linsys}) to have non-trivial solutions for the coefficients $\tilde{a}_o$ and $a_i$ is that $\det {\cal M}(iq R)=0$. By writing $\det {\cal M}= {\cal C}_o^T \gamma^1 {\cal C}_i$, where the superscript $T$ denotes the matrix transpose, we obtain the secular equation for the bound state energies:
\begin{equation}\label{secular}
\left(
\begin{array}{cccc}
H^{(1)}_l(iq\, R)& & &\frac{-q}{E+m} \,H^{(1)}_{l+1}(iq R)
\end{array}
\right)\,\gamma^1\, \Lambda\,\left(
\begin{array}{c}
J_l(iq R)\\
\frac{-q}{E+m} \,J_{l+1}(iq R)
\end{array}
\right)= 0\, .
\end{equation}
From the above equation we observe that the phase factor $e^{i\varphi}$ of the $\Lambda$ matrix again \emph{does not   affect the energies of the eventual bound state}. In terms of the $\Lambda$ matrix parameters, the above equation reads as
\begin{eqnarray}\nonumber
&&I_l\left(q R\right) \left[a\,
   q \,K_{l+1}\left(q R\right)+c\,
   (m+E )\, K_l\left(q R\right)\right]\\ \label{secularex}
&&\quad +\,\, I_{l+1}\left(q R\right) \left[b\, (m-E)\,
   K_{l+1}\left(q R\right)+d\,
   q\, K_l\left(q R\right)\right]=0\, ,
\end{eqnarray}
where $I_l(z)$ and $K_l(z)$ represent the modified Bessel functions of the first and second kinds, respectively. Again, the above secular equation can be written in terms of the physical parameters $B$ and $A_\mu$ ($\mu=0,1,2$) using (\ref{ap})-(\ref{dp}).

\subsection{Scattering states ($|E|> m$)}

The scattering states correspond to energies $|E|> m$, such that $p=\sqrt{E^2-m^2}>0$. The outer solution, written in terms of the Bessel functions  of the first and second kinds, is given by (\ref{inout}):
\begin{equation}
\Phi_o= a_o 
\left(
\begin{array}{c}
J_l(p\, r)\\
\frac{i p}{E+m} \,J_{l+1}(p\, r)
\end{array}
\right)
+
b_o 
\left(
\begin{array}{c}
Y_l(p\, r)\\
\frac{i p}{E+m} \,Y_{l+1}(p\, r)
\end{array}
\right)\, .
\end{equation}
The asymptotic behavior of these Bessel functions, that is,
$$
J_l(p\, r)\approx \sqrt{\frac{2}{\pi p r}}\cos \left(p r-\frac{1}{4} (2 l+1)\pi \right), \qquad 
Y_l(p\, r)\approx \sqrt{\frac{2}{\pi p r}}\sin \left(p r-\frac{1}{4} (2 l+1)\pi \right),
$$
suggests the following way of writing the coefficients $a_o,b_o$, in terms of the \emph{phase shift} $\delta_l$: 
\begin{equation}\label{parps}
a_o= h \cos\delta_l, \qquad 
b_o= -h \sin\delta_l,
\end{equation}
where $h$ may be complex, but $\delta_l$ is real. Therefore,
\begin{equation}
\Phi_o= a_o 
\left(
\begin{array}{c}
J_l(p\, r)-\left(\tan \delta_l\right)\, Y_l(p\, r)\\
\frac{i p}{E+m} \left[\,J_{l+1}(p\, r) -\left(\tan\delta_l\right) \,Y_{l+1}(p\, r) \right]
\end{array}
\right)\, ,
\end{equation}
and the matching condition (\ref{Lambda}) at $r=R$ gives
\begin{equation}
\left[  
\left(\begin{array}{c}
J_l(p R)-\left(\tan \delta_l\right) Y_l(p R) \\
\frac{i p}{E+m} \left[J_{l+1}(p R) -\left(\tan\delta_l\right) Y_{l+1}(p R) \right] 
\end{array}
\right) 
-\Lambda 
\left( \begin{array}{c}
J_l(p R)\\
\frac{i p}{E+m} J_{l+1}(p R)
\end{array}
\right)
\right]
\left(\begin{array}{c}
a_o\\
a_i
\end{array}
\right)=
\left(\begin{array}{c}
0\\
0
\end{array}
\right) ,
\end{equation}
where we have represented the columns of the left-hand side matrix as column matrices. To obtain non-trivial solutions for $a_o, a_i$,  the determinant of the matrix on the left-hand side must be zero. Expressed in terms of the $\Lambda$ matrix parameters, this requirement results in the following expression for the phase shift $\delta_l$:
\begin{equation}
\tan\delta_l = \frac{p (a-d)
   J_{l+1}(pR) J_l(pR)-b (E-m)
   J_{l+1}(pR){}^2+c (m+E )
   J_l(pR){}^2}{J_l(pR)
   \left[a p Y_{l+1}(pR)+c (m+E )
   Y_l(pR)\right]- J_{l+1}(pR) \left[b (E-m)
   Y_{l+1}(pR)+dp Y_l(pR)\right]} . \label{tanps}
\end{equation}
From the above expression for the phase shift, we can calculate the \emph{Wigner time delay} \cite{WTDR}
\begin{equation}\label{wtd}
\tau_l (E) =2 \frac{\partial \delta_l}{\partial E}=\left( \frac{2}{1+\tan^2 \delta_l} \right)\,\frac{\partial }{\partial E}\tan \delta_l \, ,
\end{equation}
which gives us a scale for the time the particle spends in the inner region $r<R$. The above equations for the phase shift $\delta_l$ and the Wigner time delay $\tau_l$ do not depend on the phase factor $e^{i\varphi}$ of the matrix $\Lambda$ and again these expressions can be expressed in terms of the physical parameters using (\ref{ap})-(\ref{dp}).

%
%

\subsection{Resonances}
\label{resonances}

A very practical method for investigating resonances is to look for the energies at which the scattering states become \emph{purely outgoing states}. Of course, it is not physically possible to have only purely outgoing scattering states, and for this we need to search for complex energies. The real part of these complex energies will correspond to the resonances. Although not physically possible, purely outgoing scattering states contain important information about the system: in addition to giving the resonant energies, the inverse of the imaginary parts of these complex energies provides scales for the corresponding resonant time delays, as we will see later.

A second, more physical way to characterize the resonant energies of the scattering states is to associate them with the peaks of the Wigner time delay $\tau_l(E)$ at \eqref{wtd}. This is quite intuitive, since resonant states are associated with  larger amounts of time that the particle spends within the inner region $r<R$. We will later show, in the context of the special cases investigated, that these two criteria give the same values for the resonances.

To identify the outcoming and incoming scattering states, it is convenient to write the scattering solution in the external region in terms of the Hankel functions (the second line of equations (\ref{inout})). We can use the asymptotic behavior of the Hankel functions (\ref{asympH}), with $q=-ip$, to identify the outcoming and incoming scattering solutions as $H_l^{(1)}$ and $H_l^{(2)}$, respectively. 
Thus, the purely outcoming solutions will have only the Hankel function of the first kind $H_l^{(1)}$, and thus we put $\tilde{b}_o=0$ in (\ref{inout}). The secular equation will then be exactly the same as we obtained previously for the bound states, i.e., equation (\ref{secularex}), but now with $q=-i p=-i\sqrt{\left(E_R + i E_I\right)^ 2-m^ 2}$, and we look for solutions of that equation for the complex energy $E=E_R + i E_I$.  Obviously, all solutions for a real energy in the interval $|E|\leq m$ will be included in the set of complex solutions, and these will correspond to the bound (or quasi-bound) states.


 As we continuously change the potential strengths $B, A_0, A_r$ and $A_\theta$ (or, equivalently, the parameters of the matrix $\Lambda$), new bound states can be created (\emph{captured from} the continuum) or destroyed (\emph{emitted into} the continuum). 
 A capture (emission) generally occurs when the energy of a previous resonant (bound) state continuously crosses the (super)critical values $E=\pm m$, becoming a new bound (resonant) state. The values of the potential strengths (or  $\Lambda$ parameters) for which a capture or emission from a bound state occurs are given by the conditions (\ref{critcond}) and (\ref{supercond}) for critical ($E=m$) or supercritical ($E=-m$) states, respectively.

\section{Special cases}
\label{special}

In this section, we investigate some special cases of physical interest,  where we choose the set of $\Lambda$ parameters motivated by the physical potentials singular at $r=R$ that we want to describe, according to the relations (\ref{fp})-(\ref{dp}). For each case we systematically investigate the confinement properties of the circular potential (i.e., whether it admits bound states or is impermeable), as well as the structure of the resonances and the process of capture and emission of bound states. 
In all cases, we  proceed as follows. First we  obtain the values of the potential strengths  corresponding to the critical and supercritical states, as well as the secular equation relating these strengths to the bound states energies. Next, we consider  complex-energy solutions for purely outgoing scattering states, for various values of the potential strengths,  and graphically investigate  the structure of the resonances and the discrete set of energies associated with impermeable circular walls at $r=R$. Finally, we  investigate the Wigner time delay for the scattering solutions and compare its peaks with the resonant energies calculated using the complex-energy method. 

Next we will consider five types of contact potentials concentrated at $r=R$, namely: 
(i) a purely scalar shell, (ii) a purely electrostatic shell, (iii) a  purely magnetic shell, (iv) a ``$\delta$ shell", ad (v) a ``$\delta^\prime$ shell".

%
%

\subsection{A purely scalar shell potential}
\label{pscalarshell}

First, we consider the case of a purely scalar potential, $A_{0}=A_r=A_{\theta}=0$ and $B$ arbitrary which models an infinite ``kick" in the (anti)particle mass at $r=R$.  From (\ref{lambdafields}), the parameters of the $\Lambda$ matrix are
\begin{equation}\label{escparam}
\varphi=0,\qquad a=d=\frac{4+B^2}{4-B^2},\qquad b=c=\frac{4B}{4-B^2}\, .
\end{equation}
The above relations are invariant under the transformation $B\to\frac{4}{B}$, except for a global sign, which does not affect the analysis of the critical, supercritical, bound states, or resonances. Then, for our purposes, it is sufficient to study the behavior of the system when $B\in [-2,2]$. If $B=\pm 2$, we have an impermeable wall, according to (\ref{permcond}); in such cases, the boundary conditions at $r=R$ become $\phi_{2,i}(R)=s\, \phi_{1,i}$ and $\phi_{2,0}(R)=-s\, \phi_{1,o}$, with $s=\text{sign}(B)$.


\subsubsection{Critical, supercritical and bound states} The condition (\ref{critcond}) for critical states reads
\begin{equation}\label{critesc}
\frac{c}{a}=\frac{4 B}{4+ B^2}=-\frac{l}{m R}.
\end{equation} 
For $l=0$ there is only the free solution $B=0$. For $l>0$ the solutions are
\begin{equation}\label{Bcrit}
B_\text{crit}=-\frac{2}{l}\left(mR\pm \sqrt{m^2R^2 - l^2}  \right).
\end{equation}
Thus, a necessary condition for having critical states is  $0<l\leq m\,R$. These two solutions are related to each other by the transformation $B\to\frac{4}{B}$. On the other hand, the condition (\ref{supercond}) for supercritical states is
$$
\frac{b}{a}=\frac{4 B}{4+ B^2}=-\frac{l+1}{m R},
$$
which is  similar to the condition for critical states, changing $l\to l+1$.  We will have supercritical states only for $0\leq l\leq m R-1$, and for the  following two values of $B$
$$
B_\text{super}=-\frac{2}{l+1}\left(mR \pm   \sqrt{m^2R^2 - (l+1)^2}      \right),
$$
which are also related between themselves by the transformation $B\to \frac{4}{B}$, as they should be. 
From the previous results we observe that if for a value of $B$ we have a critical state for the angular momentum $l$, for this \emph{same} value of $B$ we will have a supercritical state for the angular momentum $l-1$.

\begin{figure}[htb]
\includegraphics[width=.55\textwidth]{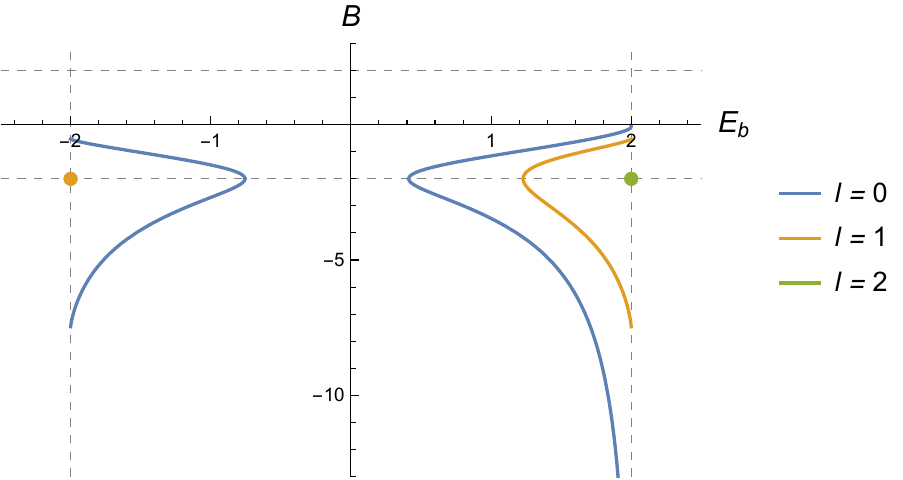}
\caption{\footnotesize \label{figboundPS} 
Graph of the energies of the bound states $E_b$ and the potential intensity $B$ for a purely scalar singular potential concentrated on the circumference $r=R$, when $m=2$ and $R=1$. Bound states can only exist for angular momentum in the range $0\leq l\leq mR$. Supercritical states ($E_b=-m$) are quasi-bound states for $l=0$. For $l=1$, the orange dot corresponds to an impenetrable circumference ($B=-2$), and $E=-m$ corresponds to the supercritical state of a particle confined in the  \emph{inner} region of the circle. Critical states are quasi-bound for $l=0,1$. For $l=2$, the green dot corresponds to an impermeable wall ($B=-2$) and $E=+m$ corresponds to the critical  state of a particle confined in the \emph{outer} region. The curves in the figure have the symmetry $B\to \frac{4}{B}$.}
\end{figure}

The secular equation (\ref{secularex}) for the bound states, with (\ref{escparam}), becomes:
\begin{equation}\label{secscalar}
    \frac{a}{c}=\frac{4+B^2}{4B}= R (E-m ) I_{l+1}\left(qR\right) K_{l+1}\left(qR\right)-R (E+m) I_l\left(qR\right) K_l\left(qR\right)\,.
\end{equation}
The relationship between the bound state energies and the strength of the pure scalar singular field $B$ is shown in Figure~\ref{figboundPS}, where it is observed that emission/absoprtion of bound states at the critical and/or supercritical energies is only allowed  for angular momenta in the range $0\leq l \leq mR$, as mentioned above. For $l>mR$, no bound states exist, regardless of the value of the scalar field strength $B$. 

 \subsubsection{Resonances} 
\begin{figure}[htb]
     \centering
   \includegraphics[width=.55\textwidth]{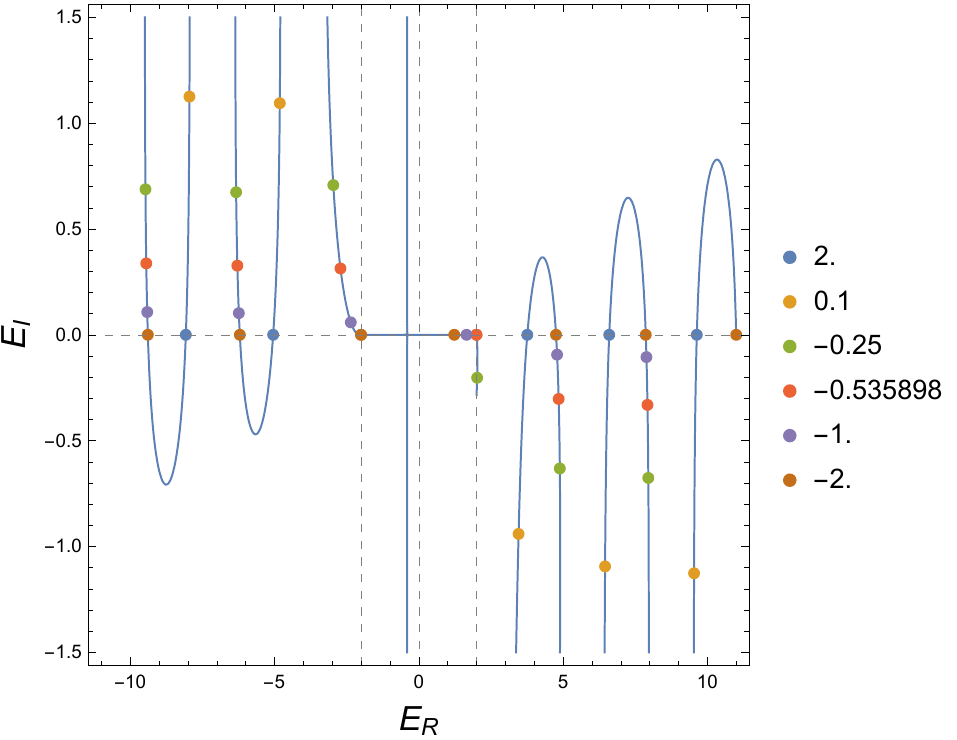} 
 \caption{ \footnotesize \label{PScalarComplex}  
The complex energy solutions of (\ref{secscalar}), corresponding to a singular pure scalar potential of intensity $B$ concentrated on the circumference $r=R$, are represented by points of different colors, for $m=2, R=1$ and $l=1$, and various values of the intensity $B$ in the interval $[-2,2]$. 
The blue curve represents the \emph{locus} of the complex solutions to the imaginary part of (\ref{secscalar}), which does not depend on the strength $B$. The values $B=\pm 2$ correspond to an impenetrable wall at $r=R$ (with boundary conditions depending on the sign of $B$), and the corresponding real energy solutions are the admissible energies for an (anti)particle confined inside the circular box (except for the energy $0<E_b<+m$), in the case $B=-2$, which corresponds to a bound state confined in the \emph{outer} region). 
As $B$ increases in absolute value the complex solutions shift towards the real axis, along the blue curve. The red points correspond to the critical value $B=B_{\text{crit}}$, at which absorption/emission of a bound state occurs at $E=+m$.}
 \end{figure}

The complex energy solutions for the secular equation (\ref{secscalar}) are shown in Figure~\ref{PScalarComplex} as colored dots, for various positive values of the scalar strength $B$ in the interval $[-2,2]$, for $l=1$, $m=2$ and $R=1$. The blue curve represent the \emph{locus} of the complex solutions to the imaginary part of equation (\ref{secscalar}), which does not depend on the strength $B$.  
For $B=\pm 2$ the circular wall becomes impenetrable (with boundary conditions depending on the sign of $B$, as  seen above), and all complex energy solutions become real, with  discrete energies being admissible for an (anti)particle confined in the inner circle, except, in the case $B=-2$, for the point (in brown in the figure) with energy $0<E_b<+E$, which corresponds to a bound state confined in the  \emph{outer} region of the circle (this state vanishes identically in the inner region, and is decaying in the outer region, as can be seen using the boundary conditions for this case).  
In the figure we observe that as $B>0$ and decreases from $2$ to $0$ the complex energies move  away from the real axis along the blue U-shaped  curves. On the other hand, as $B\in [-2,0)$ increases in absolute value, the complex solutions move towards the real axis, with the absorption of the first bound state at $E=+m$ when $B$ reaches the value $B_{crit}\in[-2,2]$ (red dots in the figure).

 \subsubsection{Wigner time delay} Equation (\ref{tanps}), in terms of the parameter $B$ of (\ref{escparam}), becomes
\begin{eqnarray*}
\tan \delta_l&=& \left\{2 \pi  B R \left[(E -m) J^2_{l+1}\left(R \sqrt{E
   ^2-m^2}\right)-(E+m ) J^2_l\left(R \sqrt{E
   ^2-m^2}\right)\right]\right\}\\
   &&\times
   \left\{ 
   B^2+2 \pi  B R \left[(E -m)
   J_{l+1}\left(R \sqrt{E ^2-m^2}\right) Y_{l+1}\left(R
   \sqrt{E ^2-m^2}\right)\right.\right.\\
   &&\left. \left. -(E+m ) J_l\left(R \sqrt{E
   ^2-m^2}\right) Y_l\left(R \sqrt{E ^2-m^2}\right)\right]+4\right\}^{-1}\, ,
\end{eqnarray*}
from which the Wigner time delay is obtained from (\ref{wtd}). 
\begin{figure}[htb]
    \centering
   \includegraphics[scale=0.6]{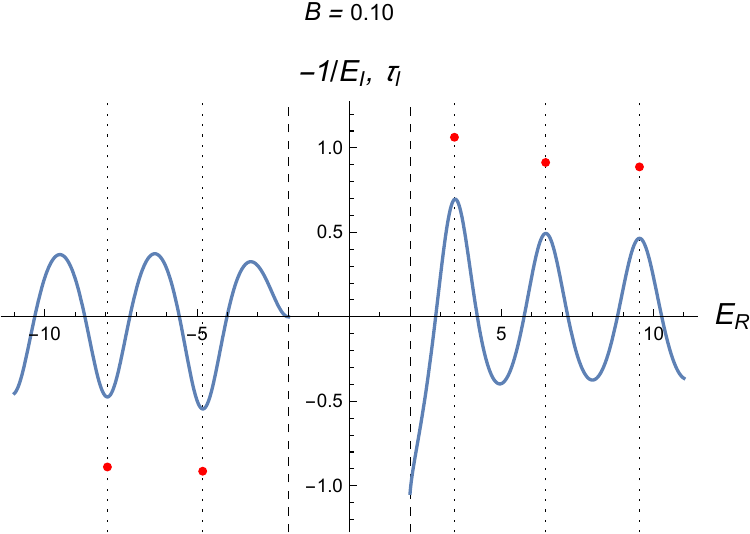} \qquad
   \includegraphics[scale=0.6]{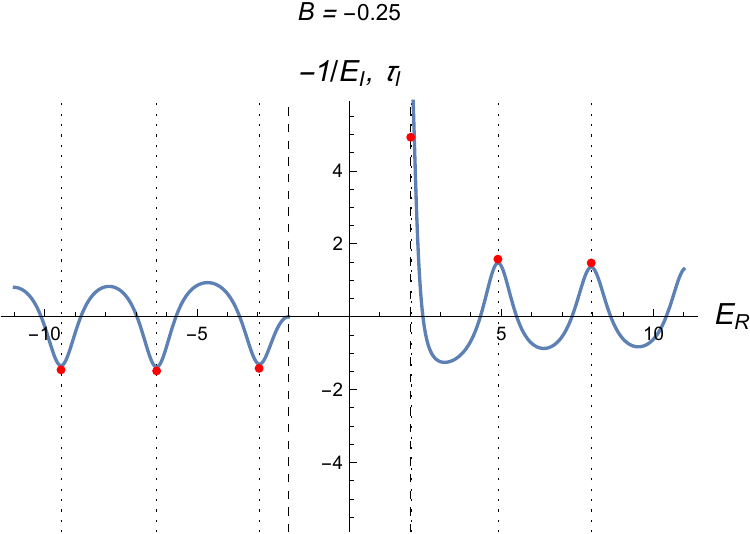} \\ [2ex]
   \includegraphics[scale=0.6]{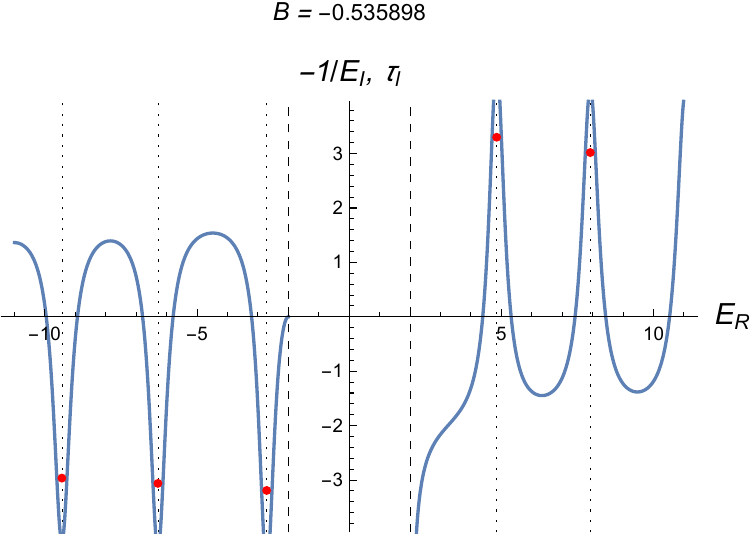}  \qquad
   \includegraphics[scale=0.6]{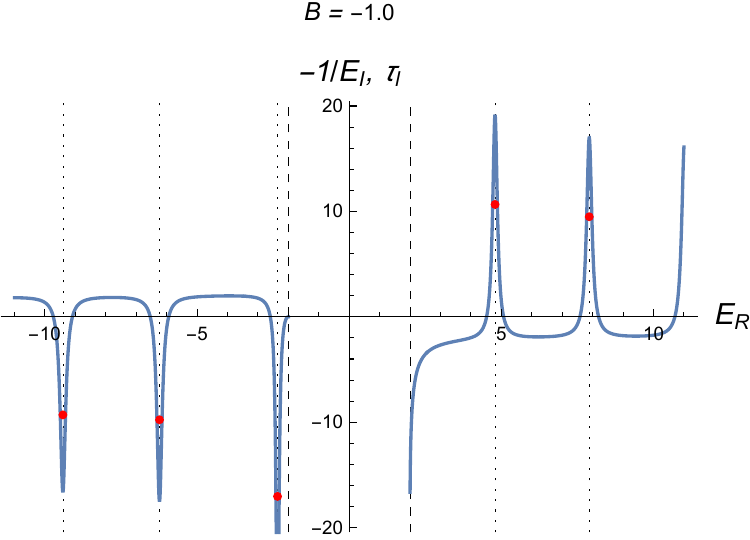}
 \caption{ \footnotesize \label{figwtdPS} 
The Wigner time delay $\tau_l$ (blue curves) for scattering states as a function of energy, for some values of the parameter $B$ from (\ref{escparam}) for a purely scalar strength ($A_0=A_r=A_\theta=0$, arbitrary $B$), and for the same parameters as in Figure~\ref{PScalarComplex}. The red points correspond to $\left(E_R,-\frac{1}{E_I}\right)$ where $E=E_R + i E_I$ are the complex energies that solve the equation (\ref{secscalar}). We note that the resonant energies $E_R$ approximate very well the sharp peaks (or valleys if $E<-m$) of the Wigner time delay, and $-1/E_I$ provide a scale for the Wigner time delay at the resonances.}
\end{figure}
Figure \ref{figwtdPS} shows this time for some values of the parameter $B$.  We observe very good agreement between the sharp peaks (or valleys, in the case of negative energies) of the Wigner time delay and the real parts of the complex energies that solve (\ref{secscalar}). These plots also confirm that the negative inverse of the imaginary parts of the complex energies provides a scale for the Wigner time delay at the resonances.
%


%
%

\subsection{A purely electrostatic shell potential}
\label{pelectrostaticshell}

A purely electrostatic shell at $r=R$ will have $B=A_r=A_\theta=0$, and an arbitrary $A_0$. From (\ref{lambdafields}), the parameters of the matrix $\Lambda$  in this case are
\begin{equation}\label{elecparam}
\varphi=0,\qquad a=d=\frac{4-A_0^2}{4+A_0^2},\qquad c=-b=\frac{4A_0}{4+A_0^2}\, .
\end{equation}
We note that the transformation $A_0\to -\frac{4}{A_0}$ converts  $a,b,c,d \to -a,-b,-c,-d$, which is equivalent to multiplying the matrix $\Lambda$  in (\ref{Lambda}) by $-1$. As already mentioned,  multiplying $\Lambda$  by a phase factor (in fact, by any constant factor) does not affect the results for bound states, resonances, or phase delay. Thus, it is sufficient to investigate those quantities for positive values of $A_0$. 
We also note that there is no value of $A_0$ for which the circular wall is impermeable, since the permeability condition (\ref{permcond}) always holds for a purely electrostatic potential. In the limit $|A_0|\to\infty$ the circular wall becomes completely transparent, since $\Lambda \to - \mathbb{1}$, which is a phase times  the matrix $\Lambda$  of the free case.\footnote{Complete transparency in the limit $A_0\to \pm \infty$ could also be concluded from the symmetry $A_0\to -\frac{4}{A_0}$.} 


\subsubsection{Critical, supercritical and bound states} 
The equation (\ref{critcond}) for critical states, written in terms of the electrostatic strength $A_0$ is
$$
- l\, A_0^2 + 4 m R\, A_0 + 4 l=0\,.
$$
For $l=0$, this equation has only the trivial solution $A_0=0$, which corresponds to the absence of interaction (free case). For $l>0$, we have two solutions,
\begin{equation}\label{critA0}
A_0=\frac{2 m R}{l}\left(1\pm \sqrt{1+\frac{l^2}{m^2 R^2}}\right)\, ,
\end{equation}
which are related to each other by the transformation $A_0\to -\frac{4}{A_0}$. Similarly, the equation (\ref{supercond}) for the supercritical states is
$$
-(l+1)\, A_0^2 -4 m R\, A_0 +4(l+1)=0\, ,
$$
whose solutions, for $l\geq 0$, are
\begin{equation}\label{superA0}
A_0=-\frac{2 m R}{l+1}\left(1\pm \sqrt{1+\frac{(l+1)^2}{m^2 R^2}}\right)\, ,
\end{equation}
that also shows the symmetry $A_0\to -\frac{4}{A_0}$.   

\begin{figure}[htb]
\includegraphics[width=.55\textwidth]{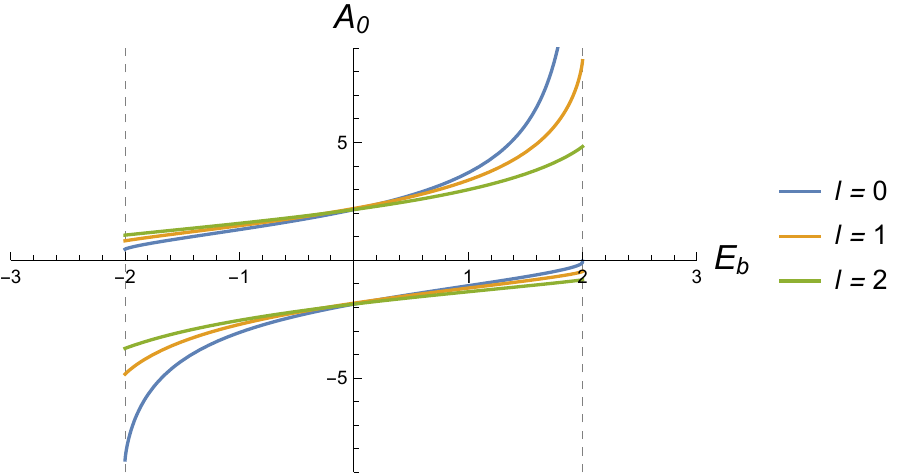}
\caption{\footnotesize\label{figboundPureElectrostatic}  
Bound state energy $E_b$ as a function of the singular electric potential strength $A_0$, for three values of the angular momentum $l$, with $m=2$ and $R=1$. A single bound state will result if the curve crosses the horizontal line defined by the value of $A_0$. 
This curve also shows the symmetry $A_0\to -\frac{4}{A_0}$, as well as the relationships between the values of $A_0$ associated to critical and supercritical states discussed in the text.}
\end{figure}

The secular equation (\ref{secularex}) for the bound states is written as 
\begin{equation}\label{secularelectrostatic}
\frac{4-A_0^2}{4A_0}= -R (E-m ) I_{l+1}\left(qR\right) K_{l+1}\left(qR\right)-R (E+m) I_l\left(qR\right) K_l\left(qR\right)\, ,
\end{equation}
expression in which we observe that any energy in the interval $(-m,m)$ will be associated to a single bound state energy for exactly two values of $A_0$.  Figure~\ref{figboundPureElectrostatic} shows the relationship between the bound state energies $E_b$ and the value of the electric potential strength $A_0$, for three values of the angular momentum $l$. 

\subsubsection{Resonances} 

\begin{figure}[htb]
    \centering
   \includegraphics[width=.55\textwidth]{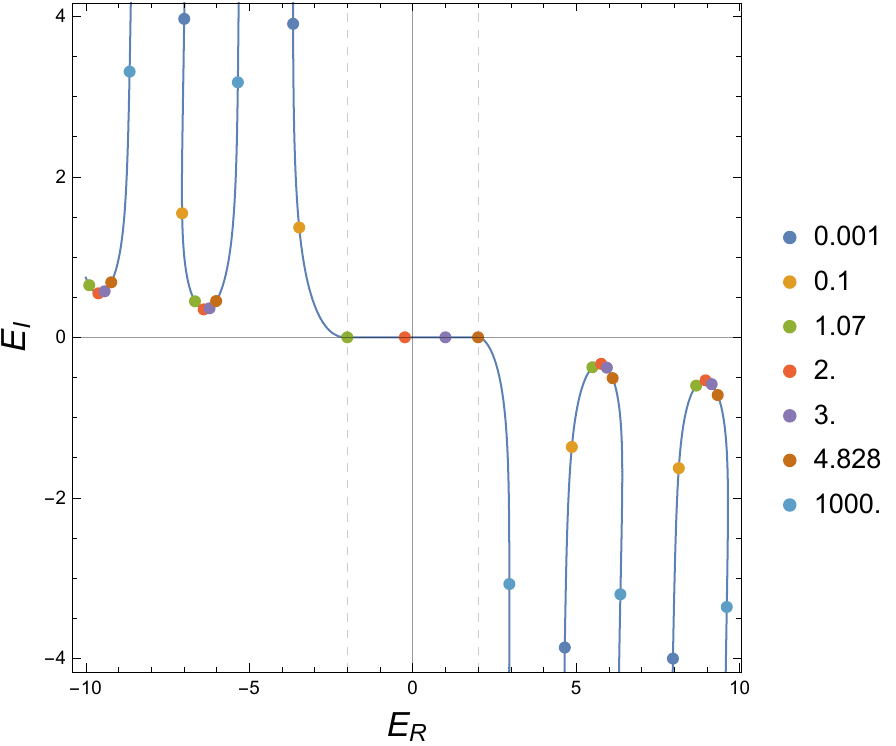} 
  \caption{ \footnotesize  
  The complex energies $E=E_{R}+i E_{I}$ that are solutions of (\ref{comel}), corresponding to purely outgoing scattering states for a pure electrostatic potential ($B=A_r=A_\theta=0$, $A_0\neq 0$), are represented by points of different colors, taking $m=2$, $R=1$ and $l=2$. The complex energies corresponding to each value of $A_0$ are shown in the same color.
The blue curve corresponds to all solutions to the real part of  (\ref{comel}), which do not depend on the value of $A_0$. As the strength $A_0$ varies from $0$ to $\infty$, the corresponding complex energies move from left to right along the blue curve. Due to the  $A_0 \to -\frac{4}{A_0}$ symmetry, the same happens as $A_0$ varies from $-\infty$ to $0$. With increasing values of $A_0$, a bound (supercritical) state is captured at $A_0\approx 1.07$ (green dot) and emitted at $A_0\approx 4.828$ (brown dot).}
  \label{PEComplex} 
 \end{figure}

We now look for  complex energy solutions to the secular equation (\ref{secularelectrostatic}), which can be rewritten in a convenient form as:
\begin{equation}\label{comel}
  (E+m)J_l(p\, R)H^{(1)}_l(p\, R)+ (E-m)J_{l+1}(p\, R)H^{(1)}_{l+1}(p\, R)=\frac{i}{2\pi R }\frac{4-A_{0}^2}{A_{0}}\,.
\end{equation}
Figure \ref{PEComplex} shows by means of colored dots the complex energies that solve (\ref{comel}), for various positive values of the electrostatic strength $A_0>0$ and $m=2, R=1, l=2$. 
The blue curve represents the solutions for the real part of (\ref{comel}), which does not depend on the potential intensity $A_0$.
As $A_0$ increases in the positive direction, the complex energy solutions move along the blue curve from left to right. When $A_0$ crosses the value (\ref{superA0}) with the minus sign ($A_0\approx 1.07$ in the figure), a bound state is captured (green dots in the figure) at $E=-m$; when $A_0$ crosses the value (\ref{critA0}) with the plus sign ($A_0\approx 4.828$ in the figure), a bound state is emitted at $E=+m$ (brown dots in the figure).

\subsubsection{Wigner time delay} 
\begin{figure}[htb]
    \centering
   \includegraphics[scale=0.6]{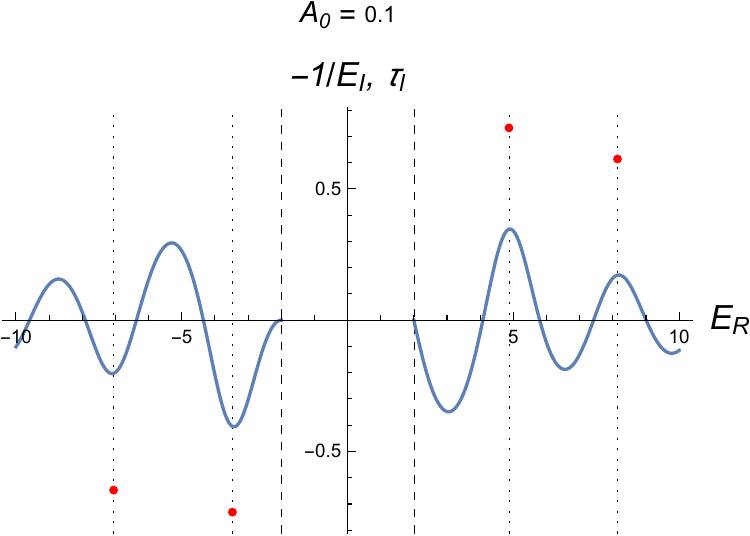}  \qquad
   \includegraphics[scale=0.6]{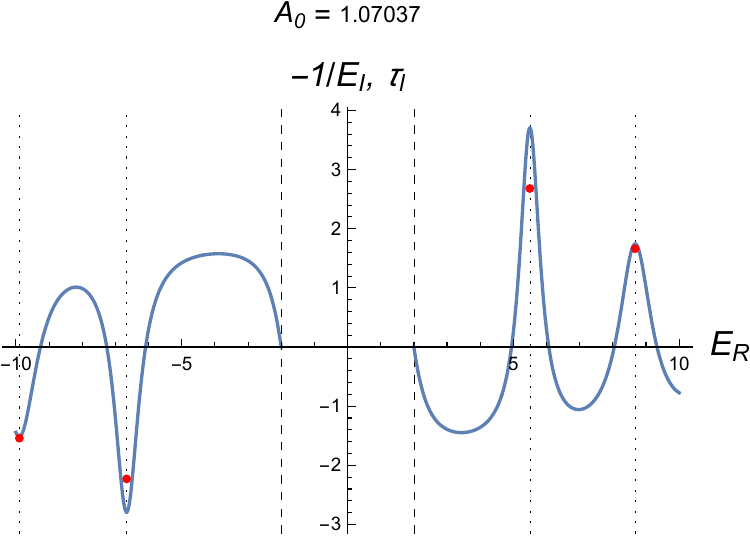} \\[2ex]
   \includegraphics[scale=0.6]{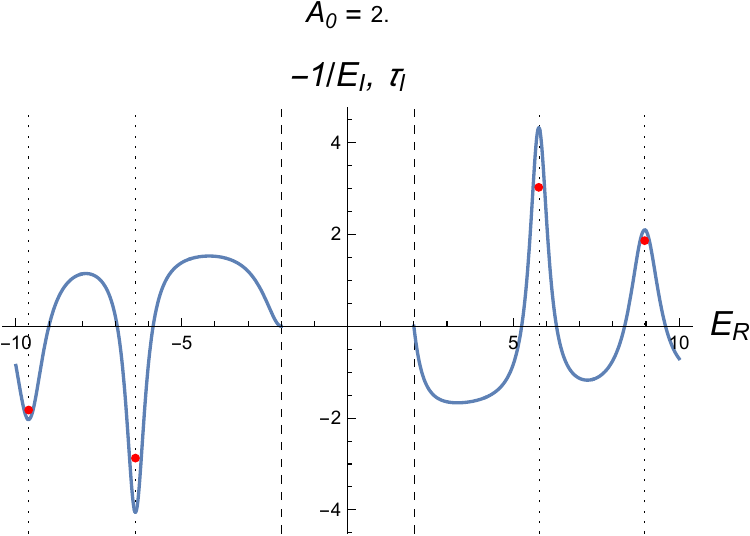}   \qquad
   \includegraphics[scale=0.6]{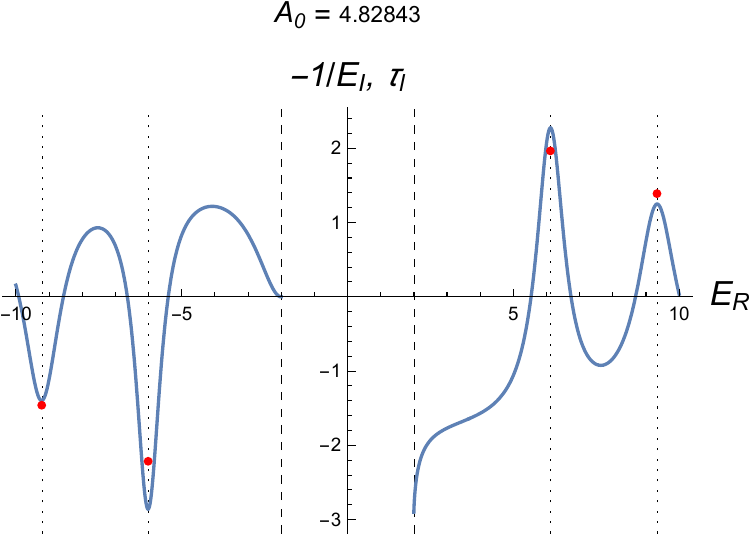}
      \caption{\footnotesize \label{figwtdPE} 
The Wigner time delay $\tau_l$ (blue curves) for scattering states as a function of energy, for some values of the strength $A_0$ of a purely electrostatic shell at $r=R$, for $m=2$, $R=1$ and $l=2$. The red points correspond to $\left(E_R,-\frac{1}{E_I}\right)$, where $E=E_R + i E_I$ are the complex energies that solve the equation (\ref{comel}). We note that the resonant energies $E_R$ approximates very well the sharp peaks (or valleys if $E<-m$) of the Wigner time delay, and $-1/E_I$ provides a scale for the Wigner time delay at the resonances.}
\end{figure}

The equation (\ref{tanps}), for the pure electrostatic case, becomes
$$
\tan \delta_l=\frac{2 \pi  A_0 R \left[(E+m) J_l\left( p R\right){}^2+(E -m) J_{l+1}\left( p R\right){}^2\right]}{A_0^2+2 \pi  A_0 R \left[(E+m ) J_l\left(p R\right) Y_l\left(p R\right)+(E -m)
   J_{l+1}\left(p R \right) Y_{l+1}\left(p R \right)\right]-4}\, ,
$$
from which we can calculate the Wigner time delay using (\ref{wtd}), shown in Figure~\ref{figwtdPE} for some values of the parameter $A_0$.  Again, we can observe a very good agreement between the sharp peaks (or valleys if the energies are negative) of the Wigner time delay and the real parts of the complex energies that solve (\ref{comel}). 

%
%

\subsection{A purely magnetic shell potential}
\label{pmagneticshell}

In this section we will consider a general pure magnetic potential ($B=A_0=0$), where $A_r$ and $A_\theta$ are not simultaneously zero. From (\ref{lambdafields}), the parameters of the $\Lambda$ matrix are
\begin{equation}\label{magparam}
\varphi=\tan^{-1} \left(\frac{4 A_r}{A_r^2+A_\theta^2 -4}\right),\quad a=\frac{1}{d}=\pm \frac{-A_r^2-(A_\theta+2)^2}{\sqrt{16A_{r}^2+(A_r^2+A_\theta^2-4)^2}},\quad b=c=0,
\end{equation}  
As mentioned above, the value of the phase $\varphi$ and the choice of the sign in (\ref{magparam}) are irrelevant to determining the energies of the critical, supercritical, bound states, and resonances, so we will only consider the negative sign in the equation above, which implies that $a=\frac{1}{d}\geq 0$. 
All the analysis of the magnetic potential will be carried out in terms of the parameter $a$, but, for clarity, let us consider some particular limits and their relationships with the physical intensities $A_r$ and $A_\theta$.
From (\ref{magparam}) we have that $a=0$ (resp. $a=\infty$) if, and only if, $A_r=0$ \emph{and} $A_\theta=-2$ (resp. $A_\theta=+2$), which are the cases for an \emph{impermeable} circular barrier, according to (\ref{permcond}). The case $a=1$ (the free case) corresponds to the limit when the strength $A_r$ or $A_\theta$ (or both) tend to infinity.


\subsubsection{Critical, supercritical and bound states} 
The condition (\ref{critcond}) for critical states in this case reads  $a\frac{l}{mR}=0$. For $l=0$ any value of $a$ admits a critical state. For $l>0$ only an impenetrable wall with $a=0$ admits a critical state, which will be a state for a particle confined in the inner region of the circular wall, as can be seen by using the boundary conditions in this case, $\phi_{2,i}(R)=\phi_{1,o}(R)=0$. On the other hand, the condition (\ref{supercond}) for supercritical states is $a\frac{l+1}{mR}=0$ and, again, only an impermeable barrier with $a=0$ will admit a supercritical state. In this case, however, the supercritical state will not correspond to an antiparticle confined in $r<R$, but in the external region $r>R$.

The secular equation (\ref{secularex}), with parameters (\ref{magparam}), becomes:
\begin{eqnarray}  \label{boundmag}
a\,I_l (q R )\,  K_{l+1} (q R )+\frac{1}{a}\, I_{l+1} (q R )\, K_l (q R )=0
\ \   \Leftrightarrow  \ \  a^2= -\frac{I_{l+1} (q R )\, K_l (q R )}{I_l (q R )\,
   K_{l+1} (q R )} .
\end{eqnarray}
The above equation has no solution for  bound state energies $|E|<m$, for any $a>0$, since $a^2$ is real and nonzero, and the right-hand side of the above equation is always negative. This result can be extended to the cases $a\to +\infty$. For the case $a=0$ we arrive at the same conclusion, explicitly considering  the boundary conditions for the impermeable barrier in this case (see above). 
In summary, a purely magnetic potential does not allow  solutions with $|E|<m$. However, as we will see later, an impenetrable wall with $a=0$ will admit a set of discrete energies with $E<-m$ or $E\geq m$  for an (anti)particle confined in the inner circle $r<R$. Similarly, an impenetrable wall with $a=+\infty$ will have states confined in the inner circle for a discrete set of energies $|E|>m$, for $l>0$, and $|E|\geq 0$, for $l=0$, as can be concluded  using the appropriate boundary conditions in this limiting case, namely $\phi_{1,i}(R)=\phi_{2,o}(R)=0$.

\subsubsection{Resonances} 
\begin{figure}[htb]
\includegraphics[width=.55\textwidth]{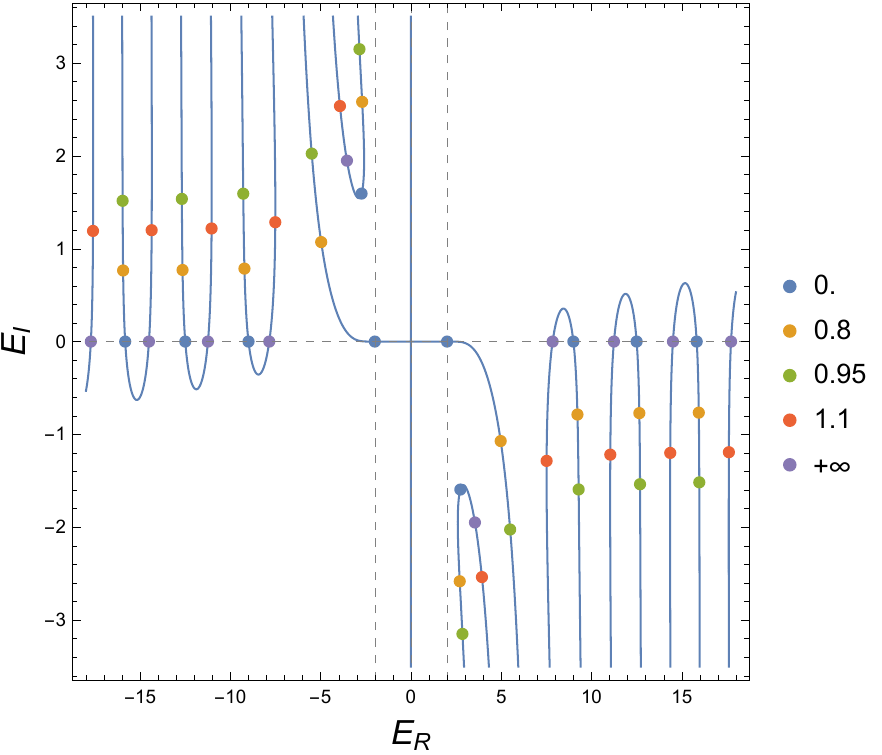}
\caption{\footnotesize\label{PMagComplex}  
Resonances in the complex plane $E=E_{R}+i E_{I}$ for particles with mass $m=2$ and angular momentum $l=4$ as a function of the parameter $a$ given in (\ref{magparam}) of a pure magnetic potential located at $R=1$. The blue solid curve correspond to the \emph{locus} of complex energies for which the imaginary part of the second equation in (\ref{boundmag}) vanishes, which does not depends on the value of the parameter $a$. The colored dots correspond to the values of complex energies that also satisfy the real part of that equation. 
As the parameter $a$ grows from $0$ to $1$ (blue, yellow and green dots), the complex energy solutions move away from the real axis, and when $a$ grows from $1$ to $+\infty$ (red and purple dots) they approach the real axis. The potential at $r=R$ turns into impenetrable walls when $a= 0$ and $a\to+\infty$ (dark blue and violet points in the figure, respectively). In both cases the purely real complex solutions correspond to (anti)particles confined in the inner wall, and both cases show two remnant resonances with $|E_R| \gtrapprox m$, which do not evolve to a confined state.}
\end{figure}

Figure \ref{PMagComplex} shows the complex energies  solving the equation (\ref{boundmag}), for various values of the parameter $a\geq 0$, and for $l=4$. As  mentioned above, the case $a=0$ (dark blue points in the figure) corresponds to an impermeable circular barrier with $A_r=0$ and $A_\theta=-2$. In this case, the purely real solutions of (\ref{boundmag}) correspond to the admissible energies for an (anti)particle confined in the inner region of the circular impenetrable wall (this set includes $E=+ m$, but not $E=-m$, which corresponds to the energy of a particle confined in the outer region). 
The two complex solutions close to $\left|E_R\right|=m$ correspond to remnant resonances. The case $a=\infty$ (purple dots), corresponding to $A_r=0, A_\theta=+2$, also produces an impenetrable circular wall (with boundary conditions on the wall different from those in the case $a=0$) and, moreover, presents a set of purely real energies corresponding to the set of discrete energies admissible for an (anti)particle confined in the inner region of the impenetrable wall. The system also presents two complex solutions, corresponding to remnant resonances close to $|E_R|= m$. 
We note (not shown in the figure) that, for lower values of the angular momentum ($l=0$ and $l=1$, with the remaining parameters fixed), such remnant complex resonances at the limits $a=0$ and $a=\infty$ do not appear. In both cases of impenetrable walls, the solutions in the outer region admits a continuum of energies $|E|>m$, as expected. For $a=0$, $E=+m$ does not belong to this continuum (it vanishes identically in the outer region), but $E=-m$ does (the solution with $E=-m$ vanishes trivially in the inner region).

\subsubsection{Wigner time delay} 
For the purely magnetic case, equation (\ref{tanps}) in terms of the parameter $a$ of (\ref{magparam}) is
$$
\tan \delta_l=\frac{\left(a^2-1\right) J_l\left(p R \right)
   J_{l+1}\left( p R \right)}{a^2 J_l\left(p R
   \right) Y_{l+1}\left(p R \right)-J_{l+1}\left(p R \right)
   Y_l\left(p R\right)}\,.
$$
\begin{figure}[htb]
    \centering
   \includegraphics[scale=0.6]{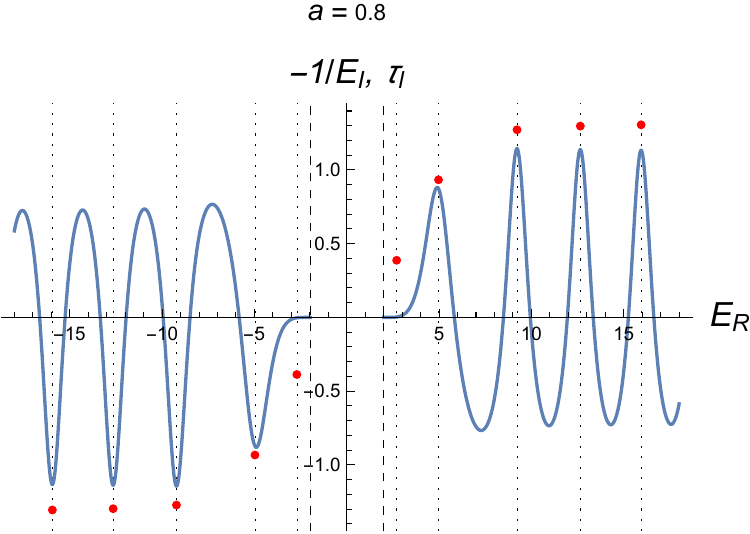} 
   \includegraphics[scale=0.6]{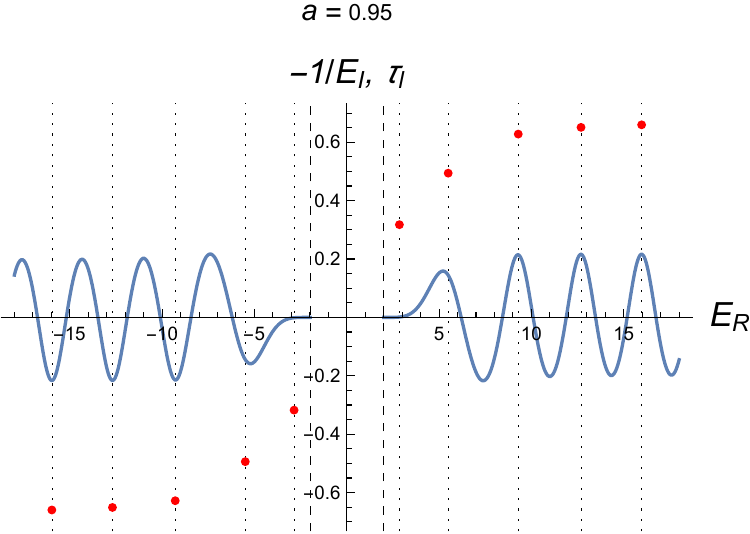} 
   \includegraphics[scale=0.6]{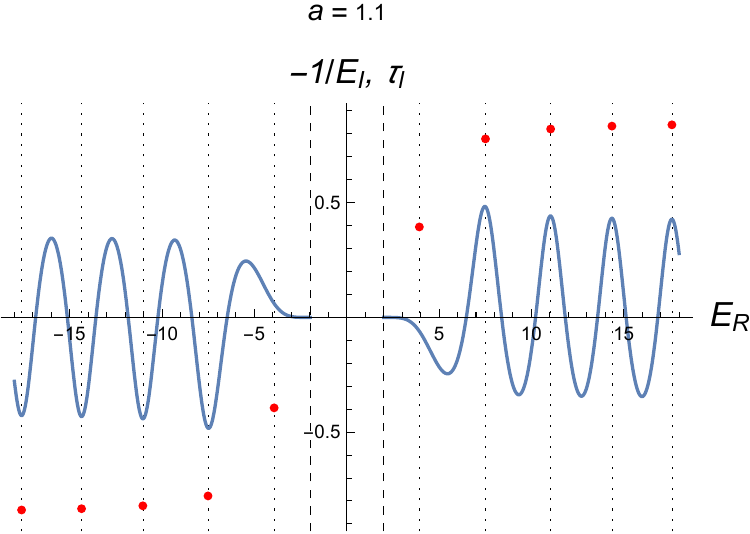} 
 \caption{ \footnotesize \label{figwtdPM} 
The Wigner time delay $\tau_l$ (blue curves) for scattering states as a function of  energy, for some values of the parameter $a$ of (\ref{magparam}) for a purely magnetic strength ($B=A_0=0, A_r, A_\theta$ arbitrary), with $m=2$, $R=1$ and $l=4$. The red dots correspond to $\left(E_R,-\frac{1}{E_I}\right)$, where $E=E_R + i E_I$ are the complex energies that solve the second equation in (\ref{boundmag}). 
We note that, for large values of $E_R$, the real part of the resonant energies, $E_R$, closely approximates  the sharp peaks (or valleys if $E<-m$) of the Wigner time delay, while $-1/E_I$ provide a scale for the Wigner time delay at such resonances. For each value of $a$, there exist two complex resonances with $|E_R|\gtrapprox m$ that do not correspond to the locations of the Wigner time delay peaks: they are the points in the upper and lower regions of Fig.~\ref{PMagComplex} whose paths along the blue curve do not evolve towards the real energies of the admitted confined state  in the limits $a\to 0,\infty$.}
\end{figure}
The Wigner time delay (\ref{wtd}) is shown in Figure~\ref{figwtdPM} for the same parameters as in Figure~\ref{PMagComplex}. 
For large values of the resonant energies (real parts of the complex energies that solve the second equation in (\ref{boundmag})), we observe a very good agreement between these energies and the locations of the sharp peaks (or valleys for negative energies) of the Wigner time delay. For these resonances, the plots also confirms that the negative inverse of the imaginary parts of the complex energies gives a scale for the Wigner time delay. In the figure, the agreement between the real and imaginary parts of the complex energies and respectively the location and magnitude of the Wigner time delay peaks is poorer for low values of $E_R$, and these relationships are absent for the two complex resonances close  to $E_R=m$. The lack of such correspondence may be understood from the fact that such resonances do not evolve to a confined state for the corresponding impenetrable barrier in the limits $a\to 0, +\infty$.
%


%
%

\subsection{A ``delta shell" potential}
\label{deltashell}

The delta shell potential corresponds to an equal mixture of the scalar and the electrostatic potentials, with the magnetic potentials vanishing (arbitrary $A_0=B$, $A_r=A_\theta=0$). We call it a ``delta shell" because the matrix $\Lambda$  in this case reduce to that associated to a delta potential in the non-relativistic limit of one-dimensional relativistic quantum mechanics (see \cite{BLM24}). From (\ref{lambdafields}), the $\Lambda$ parameter values in this case are $a=d=1$, $\varphi=b=0$, and $c=2A_0=2B$. From now on, we will characterize the strength of the delta shell potential by the value of $A_0=B$. 
%

 \subsubsection{Critical, supercritical and bound states} 
 For each $l$, condition (\ref{critcond}) gives a critical state ($E=m$) for $A_0=B= -\frac{l}{2mR}$. For $l=0$ this condition degenerates to the free case (no interaction). For $l=1$ we have a quasi-bound critical state and for $l>1$ we have a true critical bound state. On the other hand, for $l\geq 0$ equation (\ref{supercond}) has no solution for supercritical states. 
The secular equation (\ref{secularex}) for the bound state energies simplifies to 
\begin{equation}\nonumber 
I_l (q R ) \left[q \,K_{l+1} (q R )+c\,
   (m+E )\, K_l (q R )\right] + q\, I_{l+1} (q R ) K_l (q R )=0, \quad q=\sqrt{m^ 2-E^ 2},
\end{equation}
and then 
\begin{equation}\label{seculardelta}
A_0= B=-\frac{1}{2 R (E+m)\, I_l (qR ) \, K_l (qR )}  .
\end{equation}
\begin{figure}[htb]
\includegraphics[width=.55\textwidth]{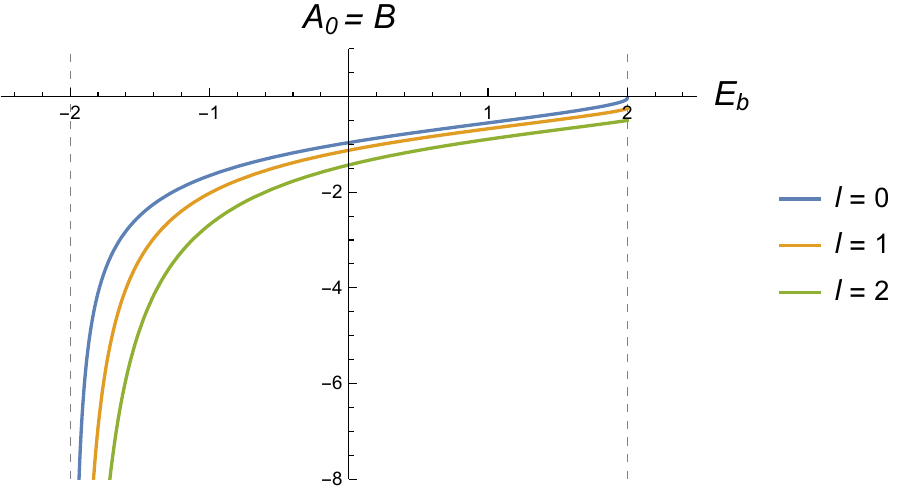}
\caption{\footnotesize\label{figdeltabound}  
Bound state energy $E_b$ for each value of the potential strengths $A_0=B$, and for three values of the angular momentum $l$. In this plot $m=2$ and $R=1$. For $l=0,1,2$ the system  capture/emit a (quasi)bound state at $E=m$ when $A_0=B=0,-\frac{1}{2}, -1$, respectively. Due to the absence of supercritical states for $l\geq 0$, in these case there are no emissions/capture of bound states at $E=-m$.}
\end{figure}
Figure \ref{figdeltabound} shows the relationship between the potential strength $A_0=B$ and the bound state energies $E_b$ for three values of the angular momentum  $l$. 
From this figure, and also from (\ref{seculardelta}), it follows that a necessary condition for having bound states is that $A_0=B<0$, since $-m<E<m$ and $I_l\left(qR\right) K_l\left(qR\right)>0$. At the value $A_0=B=-\frac{l}{2m R}$, a bound state is captured/emitted at the critical energy $E=m$. There exits a unique bound state for $A_0=B\in (-\infty,\frac{-l}{2m R}]$.  

\subsubsection{Resonances}  
We now look for the complex energy solutions to the secular equation (\ref{seculardelta}), which now becomes a complex equation that can be  conveniently rewritten as 
\begin{equation}\label{comshell}
    (E+m)J_l(p\, R)H^{(1)}_l(p\, R)=\frac{i}{\pi R A_0}\,.
\end{equation}
\begin{figure}[htb]
\includegraphics[width=.55\textwidth]{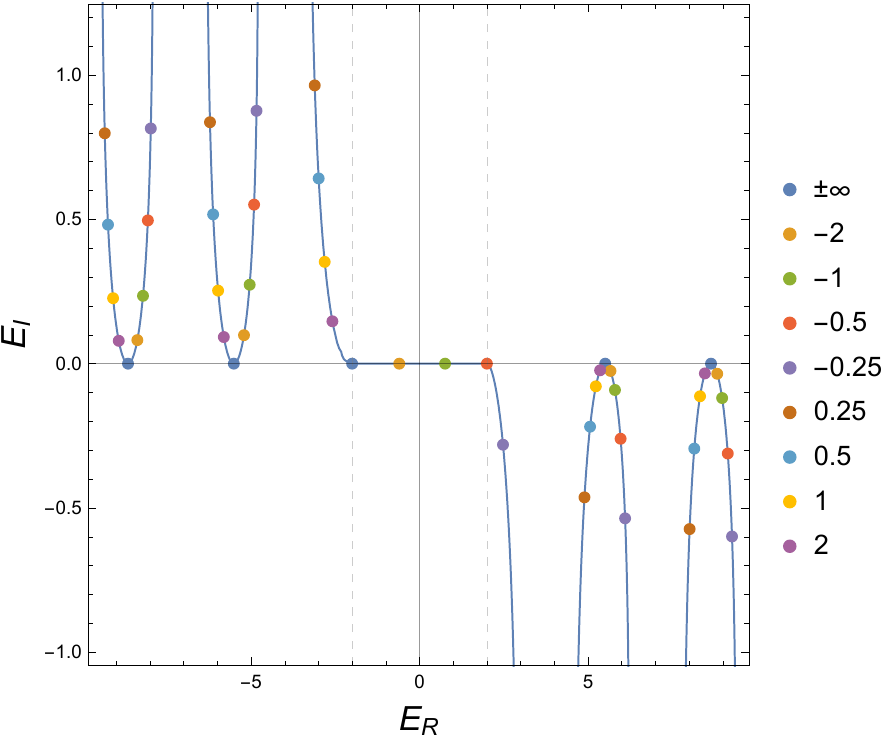}
\caption{\footnotesize\label{DSComplex}  
Complex energies $E=E_R+i E_I$ solving  (\ref{comshell}), corresponding to purely outgoing scattering states for a ``delta shell" contact potential at $r=R$ ($A_0=B$, $A_r=A_\theta=0$). In this plot, $m=2$, $R=1$ and $l=2$, and the energies are shown for several values of the potential strength $A_0=B$ (colored points). The vertical dashed lines correspond to $E_R=\pm m$. 
The real part of a complex energy corresponds to a resonant energy. In this graph, the solid blue lines correspond to the complex energies that solve the real part of (\ref{comshell}), which does not depend on the value of the potential strength $A_0$. When the potential strength crosses the value $A_0=-\frac{l}{2mR}=-0.5$ (red dots) the single bound state is emitted at $E=+m$ (or absorbed, depending on the direction of the crossing). 
The dots on the real line in the region $-m<E\leq m$ correspond to the bound state energies, while the dark blue point at $E=-m$ corresponds to an antiparticle confined in the \emph{outer} region of the impenetrable circle ($E=-m$). The other dots on the real line correspond to the discrete energies admissible for an (anti)particle confined in the \emph{inner} region of an impenetrable circumference.}
\end{figure}
Figure \ref{DSComplex} shows in blue the contour line corresponding to the complex energies that solve the real part of the equation (\ref{comshell}), which does not depend on the potential strength $A_0=B$, for $m=2, R=1$ and $l=2$. 
The colored points on that curve correspond to the complex energies that solve the whole complex equation, for various values of the potential strength $A_0=B$. The real part of the complex energies corresponds to the resonant energies. As $A_0=B$ increases in absolute value, the complex energies on the blue curve shift toward the real axis. 
When $A_0=B=-\frac{l}{2mR}$ ($=-0.5$ in the figure, denoted by red dots), the system captures/emits the unique bound state from/to the continuum. The bound state energy moves leftward along the real axis with the strength $A_0=B$ increasing in absolute value, and asymptotically approaches $E=-m$ when $A_0=B\to \pm\infty$. In these limits the circular wall becomes impermeable and all the complex energies being real. 
These energies, except $E=-m$, correspond to the discrete energies admissible for an (anti)particle confined in the inner region of the impenetrable circular wall. In this limit, there exits a supercritical energy $E=-m$, which corresponds to an antiparticle restricted to the \emph{outer} region of the impenetrable circle, as can be inferred from the associated boundary conditions in these cases, namely $\phi_{1,i}(R)=\phi_{1,o}(R)=0$ in (\ref{estado})).  

\subsubsection{The Wigner time delay} 
\begin{figure}[htb]
    \centering
   \includegraphics[scale=0.6]{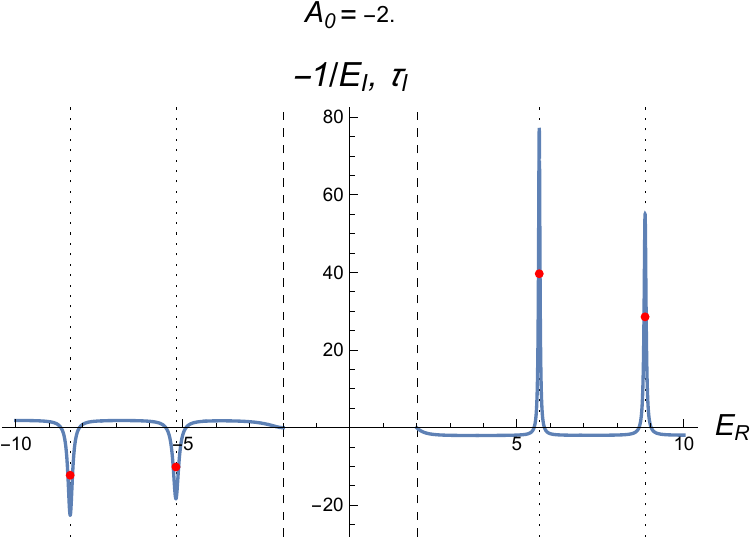} \qquad
   \includegraphics[scale=0.6]{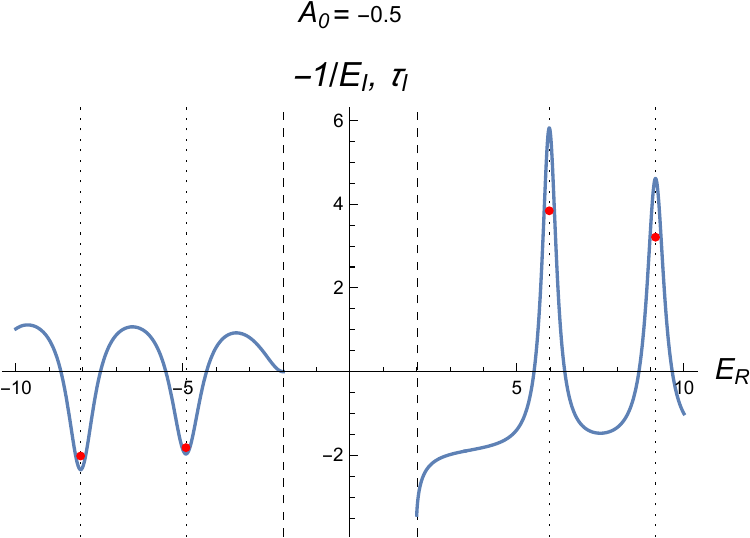} \\[2ex]
   \includegraphics[scale=0.6]{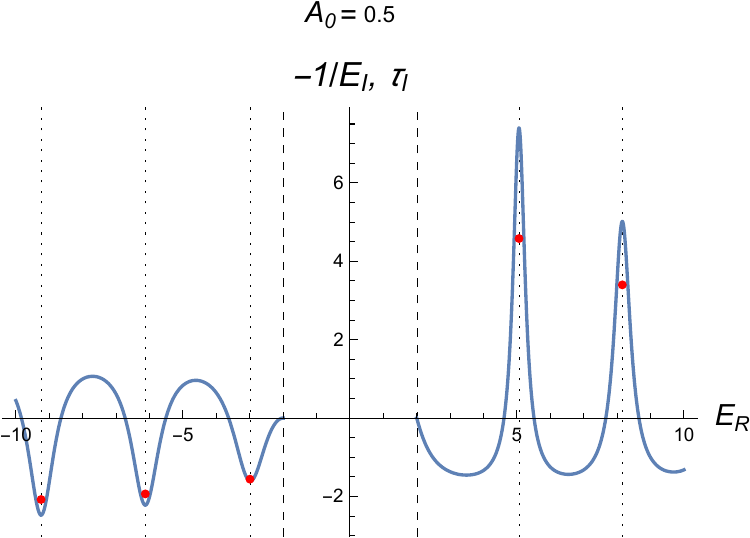} 
      \caption{ \footnotesize \label{figwtdds} 
The Wigner time delay $\tau_l$ (blue curves) for scattering states as a function of the energy, for some values of the potential strength $A_0=B$ of a delta shell contact potential at $r=R$ ($A_0=B$, $A_r=A_\theta=0$), with  $m=2$, $R=1$ and $l=2$. The red dots correspond to $\left(E_R,-\frac{1}{E_I}\right)$ where $E=E_R + i E_I$ are the complex energies that solve (\ref{comshell}). We note that the resonant energies $E_R$ approximates very well the sharp peaks (or  valleys if $E<-m$) of the Wigner time delay, and $-1/E_I$ provides a scale for the Wigner time delay at the resonances.}
\end{figure}

Now, the equation (\ref{tanps}) for the phase change becomes
\begin{equation}
\tan \delta_l = \frac{\pi R (E+m) \left[J_l\left(R p\right)\right]^2\, A_0}{\pi R (E+m)
   J_l\left(R p\right) Y_l\left(R
   p\right)  A_0-1}\, ,
\end{equation}
from which we can obtain the Wigner time delay by using (\ref{wtd}), which is shown in Figure~\ref{figwtdds}  for some values of $A_0=B$. 
We also plot the real part and the negative inverse of the imaginary part of the complex energies obtained as solutions of (\ref{comshell}). We note that the real part of the complex energies (the resonant energies) are very good approximations for the location of the Wigner time delay peaks. 
These results show that  both methods for finding resonant energies (by calculating the real part of the complex energies associated with purely outgoing scattering states or the energies related to the Wigner time delay peaks) give essentially the same results.  Moreover, the negative inverse of the imaginary part of the complex energies provides a scale for the Wigner time delay at these resonances. In the absorption/emission of a bound state at $E=+ m$ the Wigner time delay diverges, which is consistent with the  $-\frac{1}{E_I}$ scaling, since $E_I=0$ in the emission/absorption of a bound state.


%
%

\subsection{A ``$\delta^\prime$ shell" potential}
\label{deltaprimeshell}

As a final special case, we will analyze the ``$\delta^\prime$ shell", which  is an inverse combination of the scalar and electrostatic potentials, i.e., $B=-A_0, A_r=A_\theta=0$. This choice results in the following values for the $\Lambda$ matrix parameters: $a=d=1$, $c=\varphi=0$ and $b=-2A_0=2B$. It is so named because, in the non-relativistic limit of the analogous one-dimensional case, this choice of parameters results in to the so-called $\delta^\prime$  (or the non-local delta prime) interaction. 

In the following, we will show that, by means of a ``spin flip" transformation, the Dirac equation (\ref{Dirac1dsingfin}) for a ``$\delta^\prime$ shell" can be obtained from that for a ``$\delta$ shell" potential. Therefore, the results for the present case can easily be obtained from the previous one and  will not be presented here. Let us then consider the Dirac equation (\ref{Dirac1dsingfin}) for a delta shell potential, $A_0=B$, $A_r=A_\theta=0$, and a ``spin flip" transformation $\Phi\to -i \gamma^1 \Phi$. It is evident that this equation will be invariant under this transformation if also
$$
E\to -E,\qquad B\to B,\qquad A_0\to -A_0,\qquad l\to -(l+1)\, .
$$  
Therefore, the transformed equation will be the Dirac equation for a $\delta^\prime$ shell, with $B=-A_0$, $A_r=A_\theta=0$.

%
%

\section{Discussion and conclusions}
\label{epilogue}

In this work, we investigate the massive Dirac equation in the plane, with a singular interaction supported on a circumference of radius $R$. Using the radial symmetry of the interaction, we show that this problem can be reduced to a one-dimensional point interaction problem in the radial coordinate. To study this problem, we adapt the distributional approach of \cite{BLM24}, in which a physical interpretation of the parameters of point interactions in one dimension was given in terms of the strengths of four Lorentz singular point potentials. In this way, we were able to interpret the four parameters of the singular interaction supported on the circumference as the strengths of a scalar ($B$) and the three polar components ($A_0, A_r$ and $A_\theta$) of a Lorentz vector. It is worth  noting that the usual requirement of boundedness of the Dirac spinor at the origin (sometimes stated without justification) arises from the basic assumptions of the distributional approach.

After obtaining  time-independent solutions for critical, supercritical, bound, and scattering states for the most general contact potential concentrated on $r=R$, we systematically addressed five special cases of contact interactions and  their confining properties. In all these cases, we investigated how varying the potential strengths modify the structure of the critical, supercritical, bound, and resonant states. We also investigated the values (or corresponding limits) of the strengths for which the circular barrier is impenetrable. 
The resonant energies were identified using the real parts of the complex energy solutions for purely outgoing scattering conditions, and subsequently  compared to the location of the Wigner time delay peaks, with excellent overall agreement. Our results also showed that, when the real parts of the complex energies coincide with the location of the Wigner time peaks, the negative inverse of the imaginary parts provides a scale for the intensities of these peaks. 
From the permeability condition (\ref{permcond}), we observed that only potentials with strengths $A_r=0$ and a non-zero $B$ or $A_\theta$ can produce an impenetrable barrier on the circumference. For example, we saw that a purely electrostatic contact barrier cannot completely confine  an (anti)particle either in the inner or outer circle. This result is consistent with that of reference \cite{MSt87} for the one-dimensional case, and recalls Klein's paradox for point potentials.  In the following, we briefly summarize the main results for the five special cases considered in this article.

A purely scalar shell (arbitrary $B$, $A_0=A_r=A_\theta=0$),  modeling an infinite ``kick" in the (anti)particle mass at $r=R$, displays a rich structure. The ensemble of bound states and resonances has the symmetry $B\to \frac{4}{B}$, implying that an infinite strength $B$ will be similar to the free case. This potential can produce an impermeable wall at $r=R$ if the scalar strength is $B=\pm 2$ and can admit none, one, or two bound states (for $l>m R, l=m R-1$ and $l< mR-1$, respectively), for suitable negative values of the scalar strength $B$. The larger the mass, the more values of  angular momentum will admit bound states, which may be absorbed/emitted at both the critical and supercritical energies, depending on the angular momentum $l$. For any value of $B$, there exits a discrete set of resonances close to the discrete energies admitted for (anti)particles completely confined by an impenetrable circular wall. The energy location of the Wigner time delay peaks agrees very well  with the real part of the complex resonances.

A purely electrostatic shell (arbitrary $A_0$, $B=A_r=A_\theta=0$) is never impermeable, for any finite or infinite value of  strength $A_0$. For positive or negative $A_0$ and in suitable ranges, it admits a single bound state that can be absorbed/emitted at both critical and supercritical energies $E=\pm m$. The potential also exhibits a discrete set of resonances for any value of  strength $A_0$. The real part of the complex resonances also shows a very good agreement with the Wigner time delay peaks.

A purely magnetic shell (arbitrary $A_r$ and $A_\theta$  and $B=A_0=0$) does not admit bound states $|E|<m$, but it can be impenetrable for suitable combinations of finite values of the electromagnetic strengths, namely $A_0=0$ and $A_\theta=\pm 2$. 
This singular potential can therefore completely confine an (anti)particle  in the inner region of the circular wall at $r=R$, with a discrete set of admissible energies  depending on the value of $A_\theta$ ($=\pm 2$). This potential also allows  a discrete set of resonances for arbitrary choices of the electromagnetic strengths, with energies very close to the discrete energies of the impenetrable cases. It is an interesting result that for large values of the angular momentum there are still complex resonances in the limit of impenetrable walls, which do not correspond to the location of any peaks in the Wigner time delay. This mismatch may be due to the fact that these resonances do not converge to a confined state within this limit. For high values of the resonant energies, there is a good agreement between the real part of the complex energies and the locations of the Wigner time delay peaks.

Finally, we considered a situation with two ``mixed" potentials, a ``$\delta$ shell" potential (arbitrary $A_0=B$, $A_r=A_\theta=0$) and a ``$\delta^\prime$ shell" (arbitrary $A_0=-B$, $A_r=A_\theta=0$). These two cases are related  to each other by a ``spin flip" transformation, and we detail only the ``delta shell" case. The impermeability condition is met only in the limits $A_0=B\to\pm\infty$, and for both cases the system admits the same discrete set of energies for a particle confined in the inner region of the impenetrable circular wall. For finite values of the strength $A_0=B$ this potential admits a single bound state for sufficiently large and negative values of this parameter. It also admits a discrete set of resonances for any value of the strength $A_0=B$, with energies close to those corresponding to the impenetrable case.  The single bound state is captured/emitted only at the critical energy $E=+m$. The results obtained for Wigner time delay are similar to those analyzed in the previous cases.

Future applications of the contact models presented here could include the analysis of the Aharonov-Bohm effect with a singular magnetic field concentrated on a circumference. This investigation may be interesting due to the nontrivial characteristics  of singular  electromagnetic interactions, both in one and two dimensions. It also seems necessary and interesting to consider the natural generalization of these models to three dimensions, with singular potentials concentrated on a spherical surface. Finally, we believe it is worth emphasizing that our two-dimensional models presented here can describe phenomena in Dirac materials, as mentioned in the Introduction, where the Fermi velocity at the surface acts like the speed of light, and may therefore be accessible, at least in principle, to perform experiments involving ordinary velocities and short-range potentials approximately localized on a circle.

\section*{Acknowledgments}
This work is supported by Q-CAYLE (PRTRC17.11), funded by the European Union-Next Generation UE/MICIU/Plan de Recuperacion, Transformacion y Resiliencia/Junta de Castilla y Leon, and  by project PID2023-148409NB-I00, funded by MICIU/AEI/10.13039/ 501100011033. 
We also acknowledge the financial support of Castilla y León  Department of Education  and the FEDER Funds (CLU-2023-1-05).
JTL thanks Profs. M. Gadella and LMN for their warm hospitality and financial support during two short visits to the  University of Valladolid, and also the State University of Ponta Grossa for partial support related to these visits.     


%


\end{document}